\documentclass[fleqn,usenatbib]{mnras}
\usepackage{amsmath}
\usepackage{txfonts}

\usepackage[T1]{fontenc}
\usepackage{ae,aecompl}

\usepackage{graphicx}	% Including figure files
\usepackage{amssymb}	% Extra maths symbols

\newcommand {\rd}{{\rm d}}
\newcommand{\bq}{{\mathbf q}}
\newcommand{\br}{{\bf r}}
\newcommand{\bx}{{\bf x}}
\newcommand{\by}{{\bf y}}
\newcommand{\bp}{{\bf p}}
\newcommand{\bk}{{\bf k}}
\newcommand{\bg}{{\mathbf s}}
\newcommand{\mA}{\boldsymbol{\mathsf{A}}}
\newcommand{\mB}{\boldsymbol{\mathsf{B}}}
\newcommand{\mC}{\boldsymbol{\mathsf{C}}}
\newcommand{\mD}{\boldsymbol{\mathsf{D}}}
\newcommand{\mE}{\boldsymbol{\mathsf{E}}}
\newcommand{\mF}{\boldsymbol{\mathsf{F}}}
\newcommand{\mG}{\boldsymbol{\mathsf{G}}}
\newcommand{\mM}{\boldsymbol{\mathsf{M}}}
\newcommand{\mH}{\boldsymbol{\mathsf{H}}}
\newcommand{\mI}{\boldsymbol{\mathsf{I}}}
\newcommand{\mT}{\boldsymbol{\mathsf{T}}}

\newcommand{\dc}{\delta_{\rm c}}
\newcommand{\du}{\delta_{\rm r}}
\newcommand{\dv}{\bar{\delta}}%_{\cal V}}
\title[Constrained simulations]{Constrained simulations and excursion sets: understanding the risks and benefits of `genetically
modified' haloes}
\author[C. Porciani]{
Cristiano Porciani,$^{1}$\thanks{E-mail: porciani@astro.uni-bonn.de}
\\
$^{1}$ Argelander Institute for Astronomy, University of Bonn, Auf dem H\"ugel 71, D-53121, Bonn, Germany\\
}

\date{Accepted XXX. Received YYY; in original form ZZZ}

\pubyear{2016}

\begin{document}
\label{firstpage}
\pagerange{\pageref{firstpage}--\pageref{lastpage}}
\maketitle

% Abstract of the paper
\begin{abstract}
Constrained realisations of Gaussian random fields are used in cosmology to
design special initial conditions for numerical simulations. We review 
this approach and its application to density peaks
providing several worked-out examples.
We then 
critically discuss the recent proposal to use constrained realisations 
to modify the linear density field
within and around the Lagrangian patches that form dark-matter haloes.
The ambitious concept is
to forge `genetically modified' haloes with some desired properties after the non-linear evolution.
We demonstrate that the original implementation of this method
is not exact but approximate
because it tacitly assumes
that protohaloes sample a set of random points with a fixed mean overdensity. 
We show that carrying out a full genetic modification
is a formidable and daunting 
task requiring a mathematical understanding
of what determines the biased locations of protohaloes in the linear density field. 
We discuss 
approximate solutions based on educated guesses regarding the nature 
of protohaloes. 
We illustrate how the excursion-set method can be adapted to predict
the non-linear evolution of the modified patches and thus fine tune
the constraints that are necessary to obtain preselected halo properties.
This technique allows us to explore the freedom around the
original algorithm for genetic modification. 
We find that the quantity which is most sensitive to changes 
is the halo mass-accretion rate at the mass scale on which the constraints are set. 
Finally we discuss constraints based on the protohalo angular momenta.
\end{abstract}


%%%%%%%%%%%%%%%%%%%%%%%%%%%%%%%%%%%%%%%%%%%%%%%%%%

%%%%%%%%%%%%%%%%% BODY OF PAPER %%%%%%%%%%%%%%%%%%

\section{Introduction}
\citet[][hereafter HR]{HR} presented a fast technique to build constrained realisations of Gaussian random fields. This method is exact and applies as long as the constraints can be expressed in terms of linear functionals of the random field.
The algorithm has been widely used to 
generate `special' initial conditions for numerical simulations of structure
formation,
either by requiring the presence of uncommon features like high-density peaks \citep[e.g.][]{VB, RD06} 
or by imposing sets of observational constraints to
reproduce the large-scale properties of the local universe \citep[][and references therein]{GH, Sorce16}.

Recently, \citet[][hereafter RPP]{RPP} applied the HR algorithm to modify the initial conditions within the Lagrangian patches that form 
dark-matter haloes in  
numerical simulations (protohaloes).
The basic idea is to alter the linear density field in a controlled way so that to produce `genetically modified' haloes (or, possibly, even galaxies) with some desired properties (e.g. the final mass or the merging history). Although the concept is intriguing, its practical 
implementation is problematic due to the complexity of characterising the
statistical properties of protohaloes. This was already realised by \cite{MB} who considered (and then abandoned) the idea of pursuing a similar approach (see their Appendix A)
in order to build analytical models aimed at explaining the origin of the seemingly universal halo mass-density profiles.

This paper digs deeper into the matter.
In Section \ref{const}, we review the theory of constrained random fields and provide several examples of increasing complexity.
These are intended to guide the less experienced reader through the topic but also set the notation and 
provide the mathematical background to understand the rest of the paper.
Some of the examples we give are unprecedented and form the basis for new applications.
In Section \ref{gmh},
we demonstrate that the original execution of the genetic-modification idea by RPP
is approximate because it 
suffers from 
the implicit assumption that protohaloes sample a set of random points with
a fixed mean overdensity.
We show that an exact implementation of genetic modification
requires a mathematical understanding of the process of halo formation
and in particular of the physics that sets the locations of protohaloes
in the linear density field.
Using toy models rooted on the idea that protohaloes might be associated with local maxima of the smoothed density field,
we explore the degrees of freedom of genetic modification and clarify the meaning of probability of a constraint.
Our results suggest new ways to enforce constraints within protohaloes. 
In Section \ref{secmah}, we illustrate how the excursion-set method \citep[e.g.][]{BCEK, Zentner} 
can be used to predict the accretion history and the final mass of the genetically modified haloes.
This provides us with a tool to calibrate the constraints to set in order to produce a given growth history.
We also use this method to estimate the size of the deviations in the assembly history of the haloes from the solution presented in RPP.
We find that the quantity which is most affected  
is the mass-accretion rate 
at the mass scale of the constraints.
Finally, in Section \ref{am}, we discuss how to set constraints based on the angular momentum of the haloes and, in Section \ref{con},
we conclude.

\section{Theory}
\label{const}
\subsection{Conditional expectations for normal deviates}
Let ${\bf X}$  be a multivariate normal vector with expectation $E[{\bf X}]=\mathbf{m}$ and covariance matrix $\mC$.
Let us partition ${\bf X}$ into two subsets $\{{\bf Y},{\bf Z}\}$ so that $\mathbf{m}=\{\mathbf{m}_Y,\mathbf{m}_Z \}$ and write
\begin{equation}
\mC=\left(\begin{array}{cc} \mC_{YY}   & \mC_{YZ} \\
\mC_{ZY} & \mC_{ZZ} \end{array} \right)\;. 
\end{equation}
It is a classic result of probability theory that
the conditional distribution of ${\bf Y}$ given ${\bf Z}={\bf a}$ is normal with expectation 
\begin{equation}
\mathbf{m}^{\rm (c)}_Y=E[{\bf Y}|{\bf Z}={\bf a}]=
\mathbf{m}_Y+\mC_{YZ}\,\mC_{ZZ}^{-1} ({\bf a}-\mathbf{m}_Z)\;,
\label{gcondmean}
\end{equation}
and covariance matrix
\begin{equation}
\mC^{(\rm c)}_{YY}=\mC_{YY}-\mC_{YZ}\,\mC_{ZZ}^{-1}\,\mC_{ZY}\;.
\label{gcondvar}
\end{equation}
Note that the conditional covariance matrix 
$\mC^{\rm (c)}_{YY}$ does not depend on the vector ${\bf a}$. 
This property is key to building constrained realisations of Gaussian random fields (see Section \ref{constrained}).
In particular, if ${\bf Z}$ is unidimensional, the relations above reduce to:
\begin{eqnarray}
\mathbf{m}^{\rm (c)}_Y\!\!\!\!&=&\!\!\!\!\mathbf{m}_Y+\frac{\mC_{YZ}}{C_{ZZ}} (a-m_Z)\;,\label{mean1}\\
\mC^{\rm (c)}_{YY}\!\!\!\!&=&\!\!\!\!\mC_{YY}-\frac{\mC_{YZ}\,\mC_{ZY}}{C_{ZZ}} \label{variance1}\;.
\end{eqnarray}

\subsection{Constrained Gaussian random fields}
\label{constrained}
Let us consider a real-valued, stationary, Gaussian random field\footnote{To simplify the notation we will not distinguish between a finite sampling of the field in $N^3$ points (with $N\in \mathbb{N}$) forming a regular lattice (as used in
numerical simulations) 
and the continuum limit. The formal passage of letting $N\to \infty$
is discussed in \citet{Bert87} and \citet{VB}.} $\delta(\bq)$ ($\bq \in \mathbb{R}^3$) with
expectation $\langle \delta(\bq)\rangle=\mu(\bq)$. 
Let $F: \delta \to F[\delta]\in \mathbb{R}$ be a linear functional of the field that
can be generally written as %a convolution
$F[\delta]=\int h(\bq)\,\delta(\bq)\,{\rm d}^3q$
where $h$ denotes a (tempered) distribution on $\bq$-space.
It follows from Eq. (\ref{mean1}) that
the (location-dependent) mean of the field $\delta$ subject to the constraint $F[\delta]=f$ is
\begin{eqnarray}
\label{meanHR}
{\mu}^{\rm (c)}_f(\bq)\!\!\!\!\!&\equiv&\!\!\!\!\! \langle \delta(\bq) | F[\delta]=f \rangle\\&=&\!\!\!\!\!\mu(\bq)+
\frac{\langle [\delta(\bq)-\mu(\bq)] (F[\delta]-\langle F[\delta] \rangle)\rangle}{\langle (F[\delta]-\langle F[\delta]\rangle)^2 \rangle}\,(f-\langle F[\delta] \rangle)\;. \nonumber
\end{eqnarray}
Note that the symbol $\langle \dots \rangle$ denotes averages taken over all the possible realisations of the random field $\delta$ while 
 $\langle \dots | F[\delta]=f\rangle$ indicates the expected value over a restricted ensemble: only those  realisations in which $F[\delta]=f$ are considered.
Eq. (\ref{meanHR}) implies that, for each functional $F$, 
the conditional mean field ${\mu}^{\rm (c)}(\bq)$ can be written in terms of the power spectrum and the expectation of the unconstrained random field (see Section \ref{subsexamples} for further details).
Similarly, from Eq. (\ref{variance1}) we derive that the (location-dependent) variance of the constrained field
around the mean field is
\begin{eqnarray}
\label{varHR}
\Sigma^{\rm (c)}_f \!\!\!\!\!&\equiv&\!\!\!\!\!
\langle [\delta(\bq)-{\mu}^{\rm(c)}_f(\bq)]^2 | F[\delta]=f \rangle \\ &=&\!\!\!\!\!\langle [\delta(\bq)-\mu(\bq)]^2\rangle -\frac{\langle [\delta(\bq)-\mu(\bq)]( F[\delta]-\langle F[\delta]\rangle)\rangle^2}{\langle (F[\delta]-\langle F[\delta]\rangle)^2\rangle}\;.\nonumber
\end{eqnarray}

Starting from these classical results, HR developed an efficient algorithm to build a numerical realisation $\dc(\bq)$ of a Gaussian random field that satisfies the linear constraint $F[\delta]=f_{\rm c}$.
The input is an unconstrained realisation $\du$ of the random field for which it happens to be that $F[\du]=f_{\rm r}$.
This configuration can be interpreted as a specific realisation that satisfies the constraint $F[\delta]=f_{\rm r}$. Therefore one can write
$\du(\bq)={\mu}^{\rm (c)}_{f_{\rm r}}(\bq)+\epsilon(\bq)$ with $\epsilon(\bq)$ the (zero-mean) residual field with respect to the conditional mean field. Since the variance (and thus the whole probability density) of the residuals does not depend on the value of $F[\delta]$, the same $\epsilon(\bq)$ can be used to build the constrained realisation by simply adding the appropriate mean field to it: $\dc(\bq)={\mu}^{\rm (c)}_{f_{\rm c}}(\bq)+\epsilon(\bq)=\du(\bq)+{\mu}^{\rm (c)}_{f_{\rm c}}(\bq)-{\mu}^{\rm (c)}_{f_{\rm r}}(\bq)$. Putting everything together, we obtain 
\begin{equation}
\dc(\bq)-\du(\bq)=\langle \delta(\bq) | F[\delta]=f_{\rm c} \rangle-\langle \delta(\bq) | F[\delta]=f_{\rm r} \rangle\;,
\label{HRsinglewithmean}
\end{equation}
or, equivalently, using Eq. (\ref{meanHR})
\begin{equation}
\dc(\bq)-\du(\bq)=\frac{\langle [ \delta(\bq)-\mu(\bq)]( F[\delta]-\langle F[\delta]\rangle) \rangle}{\langle (F[\delta]-\langle F[\delta]\rangle)^2 \rangle}\,(f_{\rm c}-f_{\rm r})\;.
\label{HRsingle}
\end{equation}
In words: a suitably scaled mean-field component (i.e. a deterministic quantity proportional to the cross-correlation function 
between the functional constraint and the field) is added to $\du$ in order to construct a specific field realisation $\dc$ that satisfies the constraint $F[\delta]=f_{\rm c}$.
Note that the unconstrained realisation $\du$ is only used to generate the 
statistical noise around the conditional mean field. 

Equations (\ref{gcondmean}) and (\ref{gcondvar}) provide all the necessary information to impose an arbitrary number of (linear) constraints 
$F_i[\delta]=f_i$ with $i=1,\dots,N_{\rm c} \in \mathbb{N}$. In this case, the constrained mean field is 
\begin{eqnarray}
{\mu}^{\rm (c)}(\bq)\!\!\!\!\!&=&\!\!\!\!\!\langle \delta(\bq) | F_i[\delta]\nonumber=f_i \rangle\\&=&\!\!\!\!\!
\mu(\bq)+
\eta_i(\bq)\,A_{ij}^{-1}\,(f_j-\langle F_j[\delta] \rangle)\label{mfmulti}
\end{eqnarray}
(sums over repeated indices are implied)
where 
\begin{eqnarray}
\eta_i(\bq)\!\!\!\!\!&=&\!\!\!\!\!\langle [\delta(\bq)-\mu(\bq)] (F_i[\delta]-\langle F_i[\delta]\rangle) \rangle\nonumber \\ &=&\!\!\!\!\!
\langle \delta(\bq) F_i[\delta]\rangle-\mu(\bq)\langle F[\delta]\rangle
\label{mf2multi}
\end{eqnarray}
denotes the cross-covariance function between the field and the functional form of the $i$-th constraint, 
\begin{eqnarray}
A_{ij}\!\!\!\!\!&=&\!\!\!\!\!\langle (F_i[\delta]-\langle F_i[\delta]\rangle)\,(F_j[\delta]-\langle F_j[\delta]\rangle)\rangle\nonumber\\ &=&\!\!\!\!\!\langle F_i[\delta]\,F_j[\delta]\rangle-\langle F_i[\delta]\rangle \langle F_j[\delta]\rangle
\label{covcon}
\end{eqnarray}
is the $ij$ element of the covariance matrix
of the constraints $\mA$ and $\mA^{-1}$ is its inverse matrix.
Therefore, one finally obtains:
\begin{eqnarray}
\dc(\bq)-\du(\bq)\!\!\!\!\!&=&\!\!\!\!\!\langle \delta(\bq) |F_i[\delta]=f_{{\rm c},i} \rangle- \langle \delta(\bq) |F_i[\delta]=f_{{\rm r},i} \rangle \nonumber \\
&=&\!\!\!\!\!\eta_i(\bq)\,A^{-1}_{ij}\,(f_{{\rm c},j}-f_{{\rm r},j})\;,
\label{HRmult}
\end{eqnarray}
which from now on we will refer to as the `HR correction'.

Given the linearity of the constraints, it can be easily shown 
\citep{Bert87, VB} that the conditional probability 
\begin{equation}
{\cal P}[\delta | F_i[\delta]=f_i]=\frac{{\cal P}[\delta]}{{\cal P}(F_i[\delta]=f_i)}\;,
\label{probconst}
\end{equation}
where ${\cal P}[\delta]$ indicates the probability of an unconstrained realisation (a multivariate Gaussian in the case of finite sampling which can be written as a path integral in the continuum limit)
and
the probability of the constraints is 
${\cal P}(F_i[\delta]=f_i)\propto\exp(-\chi^2/2)$ with $\chi^2(f_i)=(f_i-\langle F_i[\delta]\rangle)\, A^{-1}_{ij}\,(f_j-\langle F_j[\delta]\rangle)$. 
This number can thus be used to quantify how likely it is that the constraints one is imposing occur.\footnote{
By diagonalising $\mA$ one can determine $N_{\rm c}$ linear combinations of the original constraints that are statistically independent.
In terms of the (orthonormal) eigenvectors ($\mathbf{e}_i$) and eigenvalues ($\lambda_i$) of $\mA$,
$\Delta \chi^2=(p_{{\rm c},i}^2-p_{{\rm r},i}^2)/\lambda_i$ 
where $p_i=\mathbf{w}\cdot \mathbf{e}_i$ denotes the projection of the vector with original components $w_j=f_j-\langle F_j[\delta]\rangle$ along the $i^{\rm th}$ eigenvector of $\mA$.  
Note that, in order to avoid the inversion of $\mA$, RPP re-wrote the HR algorithm in terms of a Gram-Schmidt process. Differently from them, we follow the original notation by HR which we find easier to interpret.}
The chance to randomly pick a realisation with values 
$F_i[\delta]=f_{{\rm c},i}$ with respect to one with $f_{{\rm r},i}$ is
${\cal P}_{\rm rel}\propto\exp(-\Delta \chi^2/2)$ with $\Delta\chi^2=\chi^2(f_{{\rm c},i})-\chi^2(f_{{\rm r},i})$. 
Since the probability distribution of the residual field $\epsilon(\bq)$ is 
independent of the constraints and the mean field depends deterministically
on them, 
this quantity essentially quantifies the relative likelihood of $\dc$ with respect to $\du$.
It also follows from Eq. (\ref{probconst}) that the conditional mean field is the most likely
realisation which is compatible with the constraints \citep{Bert87}.

\subsection{Examples}
\label{subsexamples}
In this Section we apply the theory described above to cosmological perturbations in the `Newtonian' limit.
Let $\delta(\bq)$ denote the linear mass-density fluctuations in the universe (at some fixed time after matter-radiation equality)
with expectation $\langle \delta(\bq) \rangle=0$ 
and power spectral density $\langle \tilde{\delta}(\bk) \tilde{\delta}(\bk')\rangle=(2\pi)^3\, \delta_{\rm D}(\bk+\bk')\,P(k)$ (where $\tilde{\delta}(\bk)=\int \delta(\bq)\exp{(i \bk\cdot\bq)}\,\rd^3q$ is the Fourier transform of the density field,
$\delta_{\rm D}(\bx)$ is the Dirac-delta distribution in three dimensions, and the random field is assumed to be
stationary, i.e. statistically homogeneous and isotropic).
Constraints will be imposed averaging the field (or the result of linear operators acting on it) over space with a weighting function  $W(\bq)$ 
characterized by the Fourier transform $\widetilde{W}(\bk)$. 
\citet{CSS} have shown that several statistics of protohaloes in $N$-body simulations can be accurately described using the effective window function
\begin{equation}
\widetilde{W}(k)=3A\,\frac{\sin(kR)-kR\cos(kR)}{(kR)^3} \,\exp{\left[-\frac{B\,(kR)^2}{50} \right]}\;,
\label{efffilter}
\end{equation}
where $R$ is the characteristic protohalo radius while $A\simeq 1$ and $B\simeq 1$ are fitting parameters that slightly depend
upon the redshift of halo identification and the halo mass.
To draw plots we will use this filter. 
%with $A=0.97$ and $f=1.1$ which provide a good description of protohaloes with $\bar{\delta}\simeq 3\sigma$ collapsing at redshift $z=0$ (the exact values adopted for the free parameters do not have any impact on our conclusions). 

Following a standard procedure in the analysis of random fields \citep[][hereafter BBKS]{CLH, Vanmarcke83, BBKS}, 
we introduce the spectral moments
\begin{equation}
\sigma_n^2=\int \widetilde{W}^2(\bk)\,k^{2n}P(k)\,\frac{\rd^3k}{(2\pi)^3}\;,
\end{equation}
with $n=0, 1$ and 2.
The ratio $R_{0}=\sigma_0/\sigma_1$ gives (neglecting factors of order unity\footnote{Note that our definitions for $R_0$ and $R_{\rm pk}$ differ from those in
BBKS by a factor of $3^{1/2}$.}) the typical separation between neighbouring zero up-crossings
of the smoothed density field (more rigorously, the mean number density of the up-crossings scales as $R_0^{-3}$). Similarly, $R_{\rm pk}=\sigma_1/\sigma_2$ characterises the separation between adjacent density maxima.
Finally, to quantify the spectral bandwidth, we introduce the dimensionless parameter $\gamma=R_{\rm pk}/R_0=\sigma_1^2/(\sigma_0\,\sigma_2)$. This quantity provides a measure of `spectral narrowness' (i.e. how concentrated the power is around the dominant wavenumbers) and ranges between 0 and 1: it is 1 for a single frequency spectrum (the number of maxima and zero up-crossings coincide in a plane wave) and 0 for white noise.
Note that $\gamma$ is the Pearson correlation coefficient between $\delta$ and $\nabla^2\delta$, i.e.
$\gamma=\langle \delta(\bq) \nabla^2\delta(\bq)\rangle/\{\langle [\delta(\bq)]^2\rangle\,\langle [\nabla^2\delta(\bq)]^2\rangle\}^{1/2}$.
For adiabatic perturbations in the $\Lambda$CDM model, $\gamma$ monotonically grows from 0.45 to 0.65 when the smoothing volume
increases from the protohaloes of dwarf galaxies to those of galaxy clusters.

\subsubsection{One density constraint}
\label{1dens}
As a first example, we use the HR method to impose a constraint on the value of the (volume-averaged) mass density at a particular location.
To simplify notation, we choose a coordinate system originating from this point
and consider the linear functional
\begin{equation}
F[\delta]=\int W(\bq)\, \delta({\bq})\, \rd^3q\equiv\dv\;.
\label{eq:defconst}
\end{equation} 
Note that $\dv$ is a stochastic variable whose value changes in each realisation of $\delta(\bq)$. 
We want to generate a specific realisation $\dc(\bq)$
in which $\bar{\delta}$ assumes the particular value $\bar{\delta}_{\rm c}$. Our input will be a random realisation $\du(\bq)$ in which it happens to be that  $\bar{\delta}=\bar{\delta}_{\rm r}$.
In order to apply Eqs. (\ref{HRsinglewithmean}) and (\ref{HRsingle}) to this case, we need to evaluate some statistical properties of the variable $\bar{\delta}$. 
Averaging over the ensemble of all possible realisations, we obtain $\langle \dv \rangle=0$ and $\langle \dv^2 \rangle=\sigma_0^2$.
At the same time,
\begin{equation}
\langle \delta(\bq)\,\dv\rangle=
\int W(\bp)\, \xi(|\bq-\bp|)\, \rd^3p\equiv \bar{\xi}(\bq)\;,
\end{equation}
with $\xi(q)=\langle \delta(\bx+\bq)\,\delta(\bx)\rangle$ the autocovariance function of the field $\delta$.
In terms of the power spectrum of $\delta$:
\begin{equation}
\bar{\xi}(\bq)=\int \widetilde{W}(\bk)\,P(k)\,e^{-i\bk\cdot\bq}
\,\frac{\rd^3k}{(2\pi)^3}\;.
\label{xibarpk}
\end{equation}
Note that, in general, the function $\bar{\xi}(\bq)$ is not spherically symmetric around the origin, this happens if and only if 
$W(\bq)$ has the same symmetry.
We can now use Eq. (\ref{meanHR}) to derive
the conditional mean field
and obtain that $\langle \delta(\bq)|\bar{\delta}\rangle=\bar{\delta}\,\bar{\xi}(\bq)/\sigma_0^2$. 
Since $\delta$ is statistically homogeneous, this quantity also coincides with
the average density profile around a random point with mean overdensity $\bar{\delta}$, i.e.
$\langle \delta(\bx+\bq)|\bar{\delta}(\bx)\rangle$ \citep[as originally derived in][]{Dekel81}.
%$\bq$ here can be read as the separation vector from the random point at which $\bar{\delta}$ is measured).
%
Given all this, when the single constraint $\dv=\dv_{\rm c}$ is imposed at the origin of the coordinate system, Eq. (\ref{HRsingle}) reduces to
\begin{equation}
\label{sol1dens}
\dc(\bq)-\du(\bq)=\Delta \bar{ \delta} \,\frac{\bar{\xi}(\bq)}{\sigma_0^2}\;,
%\dc(\bq)-\du(\bq)=\frac{\Delta \bar{ \delta}}{\sigma_0^2}\,\bar{\xi}(\bq)\;,
\end{equation}
where $\Delta \bar{\delta}=\dv_{\rm c}-\dv_{\rm r}$ quantifies how much the constraint changes the mean density within the smoothing volume. 
The relative probability of $\dc(\bq)$ with respect to $\du(\bq)$ corresponds to $\Delta \chi^2=(\dv_{\rm c}^2-\dv_{\rm r}^2)/\sigma_0^2$ (note that changes need not be small to get a likely configuration, i.e. changing sign to the mean density within the constrained region gives $\Delta\chi^2=0$).

\begin{figure}
	\includegraphics[width=\columnwidth]{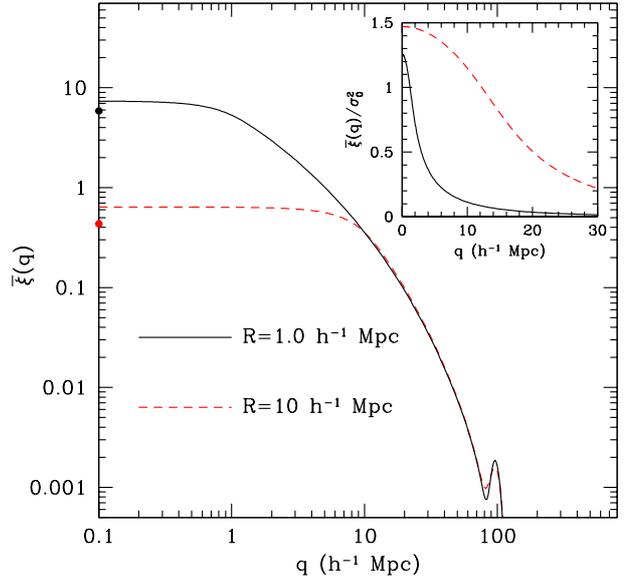}
    \caption{The curves represent the cross-covariance $\bar{\xi}(q)=\langle \delta(\bq)\,\bar{\delta}\rangle$ 
    between the linear overdensity field (extrapolated to the present time), $\delta(\bq)$, and the mean density contrast, $\bar{\delta}$, measured within a spherically symmetric region of radius $R$ centred on the origin of the coordinate system.
The feature on the right-hand side is the baryon acoustic peak.
To compute the cross-covariance we used Eqs. (\ref{efffilter}) and (\ref{xibarpk}).
 The
    small circles along the vertical axis indicate the corresponding values of the variance $\sigma_0^2=\langle \bar{\delta}^2\rangle$. The inset shows the ratio $\bar{\xi}(q)/\sigma_0^2$. This function represents the mean
    density profile around a random point with overdensity $\bar{\delta}=1$, i.e. $\langle \delta(\bq)| \bar{\delta}=1\rangle$, and
    regulates the HR correction for a single density constraint given in Eq. (\ref{sol1dens}).}
    \label{barxi}
\end{figure}

The HR correction in Eq. (\ref{sol1dens}) modifies the unconstrained field in a very specific way.
In Figure \ref{barxi} we plot the functions $ \bar{\xi}(q)$ and $\bar{\xi}(q)/\sigma_0^2$ using the Planck-2013 cosmology for a $\Lambda$CDM model and two smoothing volumes with different characteristic linear sizes $R$ (we use the window function in Eq. (\ref{efffilter}) which is spherically symmetric). 
The function $\bar{\xi}(q)$ shows a local maximum for ${\bf q}=0$.
Well within the smoothing volume,
\begin{equation}
\bar{\xi}(\bq)\simeq \bar{\xi}({\bf 0})+\frac{1}{2}\bq^{\rm T} \cdot \left[ \nabla_{\bx}\nabla_{\bx} \bar{\xi}(\bx)\right]_{\bx=\mathbf{0}}\cdot \bq+\dots
%(\bq\cdot\nabla_{\bx})^2 \left.\bar{\xi}(\bx)\right|_{\bx=\mathbf{0}}+\dots
\end{equation}
which, in the spherically symmetric case (when the traceless part of the Hessian does not contribute by symmetry), reduces to 
\begin{equation}
\bar{\xi}(q)\simeq \bar{\xi}(0)+\frac{1}{6}\nabla^2\bar{\xi}(0)\,q^2+\dots
\end{equation}
where\footnote{The Fourier integrals defining $\bar{\xi}(\mathbf{0})$ and $\nabla^2\bar{\xi}(\mathbf{0})$ are analogous to $\sigma_0^2$ and $\sigma_1^2$ but are evaluated using $\widetilde{W}(k)$ instead of its square. }
$\bar{\xi}(0)=(2\pi^2)^{-1}\int \widetilde{W}(k)\,k^2\,P(k)\,\rd k>0$ and
$\nabla^2 \bar{\xi}(0)=\overline{\nabla^2 \xi}(0)=-(2\pi^2)^{-1}\int \widetilde{W}(k)\,k^4\,P(k)\,\rd k<0$.
For $q\gg R$, instead, $\bar{\xi}(q)$
scales proportionally to the autocovariance function of $\delta$. % for $q\gg R$.
Note that
imposing a localised constraint on the size of the density fluctuations requires long-range corrections due to the slowly decreasing spatial autocorrelation of the random field $\delta$. 
If the density field has substantial power on scales smaller than $R$, then the HR correction is always subdominant with
respect to the unconstrained field (this might not be noticeable when setting the initial conditions for $N$-body simulations due to the artificial
cutoff of the power around the Nyquist frequency).  Also note that the mean density of a constrained realisation within a finite box does not vanish.

% Figure 1

\subsubsection{Two density constraints}
\label{2dens}
In some applications it is useful to set multiple constraints.
As an example we impose
two simultaneous conditions on the values of the volume-averaged mass density (defined using different smoothing volumes and denoted by the subscripts A and B) at the same spatial location (here identified as the origin of the coordinate system). 
In this case, the diagonal elements of the covariance matrix of the constraints, $\mA$,  
are $\sigma^2_{\rm A}=\langle \bar{\delta}_{\rm A}^2\rangle$ and $\sigma^2_{\rm B}=\langle \bar{\delta}_{\rm B}^2\rangle$ while the off-diagonal element is
$\zeta=\langle \bar{\delta}_{\rm A} \bar{\delta}_{\rm B}\rangle=(2\pi)^{-3}\int \widetilde{W}_{\rm A}(\bk)\,
\widetilde{W}_{\rm B}(\bk)\,P(k)\,\rd^3k$.
The appropriate HR correction straightforwardly follows from Eq. (\ref{HRmult}), 
\begin{equation}
\dc(\bq)-\du(\bq)=\alpha_{\rm A}\,\bar{\xi}_{\rm A}(\bq)+\alpha_{\rm B}\,\bar{\xi}_{\rm B}(\bq)\;,
\end{equation}
with
\begin{eqnarray}
\!\!\!\!\!\!\alpha_{\rm A}\!\!\!\!\!&=&\!\!\!\!\!(\sigma_{\rm A}^2\sigma_{\rm B}^2-\zeta^2)^{-1}\left(\sigma_{\rm B}^2\,\Delta\bar{\delta}_{\rm A}-
\zeta\,\Delta\bar{\delta}_{\rm B}\right) \\
\!\!\!\!\!\!\alpha_{\rm B}\!\!\!\!\!&=&\!\!\!\!\!(\sigma_{\rm A}^2\sigma_{\rm B}^2-\zeta^2)^{-1}\left(\sigma_{\rm A}^2\,\Delta\bar{\delta}_{\rm B}-
\zeta\,\Delta\bar{\delta}_{\rm A}\right)\;.
\end{eqnarray}
%where $\sigma^2_{\rm A}=\langle \bar{\delta}_{\rm A}^2\rangle$, $\sigma^2_{\rm B}=\langle \bar{\delta}_{\rm B}^2\rangle$ and
%$\zeta=\langle \bar{\delta}_{\rm A} \bar{\delta}_{\rm B}\rangle=(2\pi)^{-3}\int \widetilde{W}_{\rm A}(\bk)\,
%\widetilde{W}_{\rm B}(\bk)\,P(k)\,\rd^3k$.
%
The tangled structure of the solution above reflects the fact that $\bar{\delta}_{\rm A}$ and $\bar{\delta}_{\rm B}$ are correlated Gaussian variables.
The relative likelihood of $\dc$ vs. $\du$ is quantified by $\Delta\chi^2=\chi^2_{\rm c}-\chi^2_{\rm r}$ with $\chi^2=[\sigma_{\rm B}^2\bar{\delta}_{\rm A}^2+\sigma_{\rm A}^2\bar{\delta}_{\rm B}^2-2\zeta\bar{\delta}_{\rm A}\bar{\delta}_{\rm B}]/(\sigma_{\rm A}^2\sigma_{\rm B}^2-\zeta^2)$.

\subsubsection{Density and density gradient constraints}
\label{extre}
Let us now 
enforce simultaneous constraints\footnote{Constraints on the density gradient can be imposed using the 
the derivative of the Dirac-delta distribution to define the linear functional $F[\delta]$.} on the variables
$\bar{\delta}$ and $\bar{\bg}=\overline{\nabla \delta}$ at $\bq=0$
(from now on, to simplify notation, we use the same window function for all constraints but it is trivial to generalise our formulae by considering the appropriate combinations of smoothing radii to evaluate the spectral moments and $\bar{\xi}$).
In a Gaussian random field, $\langle \delta\,\nabla \delta\rangle=0$ (in general, odd derivatives are uncorrelated with even derivatives) and $\langle \bar{s}_i \,\bar{s}_j \rangle=\sigma_1^2\,\delta_{ij}/3$ (where $\delta_{ij}$ is the Kronecker symbol). The matrix $\mA$ is therefore diagonal and the cross-covariance
$\langle \delta(\bq)\,\overline{\nabla\delta}\rangle=\nabla\bar{\xi}(\bq)$.
Eq. (\ref{HRmult}) then gives
\begin{equation}
\dc(\bq)-\du(\bq)=\Delta\dv \,\frac{\bar{\xi}(\bq)}{{\sigma_{0}^2}}+
\Delta\bar{\bg}\cdot\frac{\nabla\bar{\xi}(\bq)}{{\sigma_{1}^2}}\;,
%\dc(\bq)-\du(\bq)=\frac{\Delta\dv}{\sigma_{0}^2}\,\bar{\xi}(\bq)+
%\frac{\Delta\bar{\bg}}{\sigma_{1}^2}\cdot\nabla\bar{\xi}(\bq)\;,
\label{HRdandg}
\end{equation}
with $\Delta\chi^2=(\dv_{\rm c}^2-\dv_{\rm r}^2)/\sigma_0^2+(\bar{s}^2_{\rm c}-\bar{s}^2_{\rm r})/\sigma_1^2$.
A couple of things are worth noting in the HR correction. First, setting constraints on $\overline{\nabla \delta}$ results in the appearance
of a new term proportional to $\nabla \bar{\xi}$. 
Second, contrary to $\Delta \bar{\delta}_{\rm A}$ and $\Delta \bar{\delta}_{\rm B}$ in \S \ref{2dens}, $\Delta \bar{\delta}$ and
$\Delta \bar{\bg}$ do not mix due to the fact that $\bar{\delta}$ and $\overline{\nabla \delta}$ are independent Gaussian variables. 

\subsubsection{Density and tidal-field constraints}
\label{dt}
Tides play a major role in gravitational collapse and it is certainly interesting to be able to control them in the initial conditions of numerical simulations.
We thus impose constraints on the elements of the
linear deformation tensor $\mD=\nabla\nabla\Phi$ 
with $\Phi=\nabla^{-2}\delta$ the (suitably rescaled) peculiar gravitational potential. Note that the trace of $\mD$ coincides with $\delta$ while the linear tidal tensor $\mT=\mD-(\delta/3)\mI$ (where $\mI$ denotes the identity matrix with elements $\delta_{ij}$) is
the traceless part of the deformation tensor. Considering that
$\langle \delta(\bx+\bq) \,\overline{D}_{ij}(\bx)\rangle=\partial_i\partial_j \nabla^{-2}\bar{\xi}(\bq)$ and
$\langle \overline{D}_{ij}\, \overline{D}_{\ell m}\rangle= \sigma_0^2 \,(\delta_{ij}\delta_{\ell m}+\delta_{i\ell}\delta_{jm}+\delta_{im}\delta_{\ell j})/15$,
Eq. (\ref{HRmult}) gives:
\begin{eqnarray}
\dc(\bq)-\du(\bq)=\Delta\bar{\delta}\,\frac{\bar{\xi}(\bq)}{\sigma_0^2}+\frac{15}{2}\left(\Delta \overline{T}_{ij}\partial_i\partial_j\right) \frac{\nabla^{-2}\bar{\xi}(\bq)}{\sigma_0^2}\;.
%\dc(\bq)-\du(\bq)=\frac{\Delta\bar{\delta}}{\sigma_0^2}\bar{\xi}(\bq)+\frac{15}{2\sigma_0^2}\left(\Delta \overline{T}_{ij}\partial_i\partial_j\right) \nabla^{-2}\bar{\xi}(\bq)\;.
\end{eqnarray}
In this case, $\Delta \chi^2=\chi^2_{\rm c}-\chi^2_{\rm r}$, with
\begin{eqnarray}
\chi^2\!\!\!\!\!&=&\!\!\!\!\!\frac{1}{\sigma_0^2}\left[\bar{\delta}^2+6\left(\overline{T}_{11}^2+\overline{T}_{22}^2+\overline{T}_{33}^2 \right)\right.\\
&-&\!\!\!\!\!3\left( \overline{T}_{11} \overline{T}_{22}+\overline{T}_{11}  \overline{T}_{33}
+\overline{T}_{22} \overline{T}_{33}\right)
+15\left.\left( \overline{T}_{12}^2+\overline{T}_{13}^2+\overline{T}_{23}^2\right)\right]\;,\nonumber
\end{eqnarray}
where $\overline{T}_{11}+\overline{T}_{22}+\overline{T}_{33}=0$.
\subsubsection{Adding curvature constraints}
\label{curv}
Finally,  we generalise all our previous results by imposing extra constraints on the six independent elements of the Hessian matrix $\mH=\nabla\nabla \delta$ 
in addition to controlling $\bar{\bg}$ and $\overline{\mD}$.
Since $\mH$ is made of second-order derivatives of $\delta$, it correlates with the density and the tidal fields: 
$\langle D_{ij}(\bx+\bq)\,\overline{H}_{\ell m}(\bx)\rangle=\partial_i\partial_j \partial_\ell \partial_m \nabla^{-2}\bar{\xi}(\bq)$.
At the same time, the covariance matrix of the constraints is composed of simple blocks and its inverse can be written in a compact analytic form (see Appendix \ref{inversion}). 
In fact, the only additional non-vanishing contributions to $\mA$ with respect to those discussed in \S\ref{extre} and \S\ref{dt} are
$\langle \overline{D}_{ij} \,\overline{H}_{\ell m} \rangle=\sigma_1^2 \,(\delta_{ij}\delta_{\ell m}+\delta_{i\ell}\delta_{jm}+\delta_{im}\delta_{\ell j})/15$
and $\langle \overline{H}_{ij}\, \overline{H}_{\ell m}\rangle= \sigma_2^2 \,(\delta_{ij}\delta_{\ell m}+\delta_{i\ell}\delta_{jm}+\delta_{im}\delta_{\ell j})/15$. 

After performing the matrix inversion,
we can easily derive the conditional mean field $\langle \delta(\bq) | \overline{\mD}, \bar{\bg}, \overline{\mH}\rangle$ using Eqs. (\ref{mfmulti}) and (\ref{mf2multi}). 
It is convenient to express the final results
in terms of the Laplacian $\nabla^2 \delta=H_{11}+H_{22}+H_{33}\equiv \kappa$ (which gives the sum of the principal
curvatures or, equivalently, 3 times the mean principal curvature) and of the tensor $\mC=\mH-(\kappa/3)\mI$
(the trace-free part of the Hessian matrix) which describes the orientation and the relative length of the principal axes of curvature. We thus obtain:
\begin{eqnarray}
\label{quasimain}
\langle \delta(\bq) | \overline{\mD}, \bar{\bg}, \overline{\mH}\rangle\!\!\!\!\!
&=&\!\!\!\!\!
\bigg\{\frac{1}{\sigma_0^2(1-\gamma^2)}\,\bigg[\bar{\delta}\,\left(1+R_{\rm pk}^2\,\nabla^2\right)+\bar{\kappa}\, R^2_{\rm pk}\,\left( 1+R_0^2\,\nabla^2\right) \nonumber \\
&+&\!\!\!\!\!
\overline{T}_{ij}\,\frac{15}{2}\left(\partial_i\partial_j\nabla^{-2}+R_{\rm pk}^2\, \partial_i\partial_j\right)\nonumber\\
&+&\!\!\!\!\! \overline{C}_{ij}\,R_{\rm pk}^2\,\frac{15}{2}\left(1+R_0^2\,\partial_i\partial_j \right)\bigg]
+\frac{1}{\sigma_1^2}\bar{s}_i\partial_i\bigg\}\,\bar{\xi}(\bq)\;,
\end{eqnarray}
where implicit summations run over all the nine elements of the tensors (and not over six like in Appendix \ref{inversion}).
The constrained density field is derived from Eq. (\ref{HRmult}) which, in this instance, gives
\begin{equation}
\dc(\bq)-\du(\bq)=\langle \delta(\bq) | \overline{\mD}_{\rm c}, \bar{\bg}_{\rm c}, \overline{\mH}_{\rm c}\rangle- \langle \delta(\bq) | \overline{\mD}_{\rm r}, \bar{\bg}_{\rm r}, \overline{\mH}_{\rm r}\rangle\;.
\label{main}
\end{equation}
Note that the rhs of this equation assumes
the same identical form as in Eq. (\ref{quasimain}) provided that the field variables subject to constraints are replaced
with their variations (e.g. $\bar{\delta}\to \Delta \bar{\delta}$, $\bar{\kappa}\to \Delta \bar{\kappa}$, etc.).
Once again the relative likelihood of $\dc$ vs. $\du$ is quantified by $\Delta\chi^2=\chi^2_{\rm c}-\chi^2_{\rm r}$ where, in this case, 
\begin{equation}
\label{chimain}
\chi^2=
\frac{\psi(\delta,{\mT},\delta,{\mT})}{\sigma_0^2(1-\gamma^2)}+\frac{\psi(\kappa,{\mC},\kappa,{\mC})}{\sigma_2^2(1-\gamma^2)}
-\frac{2\gamma}{1-\gamma^2}\,\frac{\psi(\delta,{\mT},\kappa,{\mC})}{\sigma_0\sigma_2}
+\frac{\bar{s}^2}{\sigma_1^2}\;,
\end{equation}
with, for instance,
\begin{eqnarray}
\psi(\delta,{\mT},\kappa,{\mC})\!\!\!\!\!&=&\!\!\!\!\! \bar{\delta} \bar{\kappa}+6\left[ \overline{T}_{11}\overline{C}_{11}+ \overline{T}_{22} \overline{C}_{22}+ \overline{T}_{33} \overline{C}_{33} \right] 
-\frac{3}{2}\left[ \overline{T}_{11} \overline{C}_{22}\right.\nonumber\\
&+&\!\!\!\!\! \left.\overline{T}_{11}  \overline{C}_{33}+\overline{T}_{22} \overline{C}_{11}
+\overline{T}_{22} \overline{C}_{33}+ \overline{T}_{33}  \overline{C}_{11}+ \overline{T}_{33} \overline{C}_{22}\right] \nonumber\\
&+&\!\!\!\!\!15\left[ \overline{T}_{12}  \overline{C}_{12}+ \overline{T}_{13} \overline{C}_{13}+ \overline{T}_{23} \overline{C}_{23}\right] \;.
\end{eqnarray}

\subsection{Setting constraints at local density maxima}
\label{hrbbks}
Some applications require setting constraints at special locations that form a point process and for which
elementary probability theory does not apply (for further details see Appendix \ref{condpeak}).
A classic example is maxima (peaks) of the smoothed linear density field which are often used
as a proxy for the location of protohaloes \citep[e.g.][BBKS]{Doroshkevich70, K84, PH}. 
A peak is a point in which $\overline{\nabla \delta}$ vanishes and $\overline{\mH}$ is negative definite. 

BBKS derived several statistical properties (e.g. the mean density and the large-scale clustering amplitude as a function of the peak characteristics) for local maxima of a random field in three
dimensions. These authors also computed the mean and variance of the mass-density profiles around peaks. 
The key element to perform these calculations is the definition of probability for $\delta$ subject
to the constraint that there is a peak at a specific location.
In general, considering only peaks with overdensity $\bar{\delta}$ and Hessian matrix $\overline{\mH}$ gives 
\begin{equation}
\langle \delta(\bq) | F[\delta]=f\rangle_{\rm pk}=
\langle \delta(\bq) | F[\delta]=f, \bar{\delta}, \bar{\bg}=0, \overline{\mH}\rangle
\label{peakcond}
\end{equation}
where the subscript pk indicates that a local density maximum is present at the origin of the coordinate system 
(see our Appendix \ref{condpeak} for a formal derivation of this equation which is not as intuitive as it might
seem).

Eq. (\ref{peakcond}) shows that conditional probabilities requiring the presence of a peak are equivalent to those
obtained imposing a set of linear constraints on $\delta$ and its spatial derivatives. It is exactly this property that makes it possible to use the HR method also for peak conditioning. 
From Eq. (\ref{peakcond}) we can write
the mean field around a peak of height $\bar{\delta}$ and Hessian matrix $\overline{\mH}$
as $\langle \delta(\bq) \rangle_{\rm pk}=\langle \delta(\bq)| \bar{\delta}, \bar{\bg}=0, \overline{\mH}\rangle$ and 
the ensemble average on the rhs can be easily evaluated using Eq. (\ref{quasimain}). We finally obtain
\begin{eqnarray}
\label{peakprofile}
\langle \delta(\bq)\rangle_{\rm pk}\!\!\!\!\!&=&\!\!\!\!\!\bigg\{\frac{1}{\sigma_0^2(1-\gamma^2)}\,\bigg[\bar{\delta}\,\left(1+R_{\rm pk}^2\,\nabla^2\right)
+\bar{\kappa}\, R^2_{\rm pk}\,\left( 1+R_0^2\,\nabla^2\right) \nonumber\\
&+&\!\!\!\!\!\overline{C}_{ij}\,R_{\rm pk}^2\,\frac{15}{2}\left(1+R_0^2\,\partial_i\partial_j \right)\bigg]\bigg\}\,\bar{\xi}(\bq)\;,
\end{eqnarray}
which coincides with Eq. (7.8) in BBKS although it is written using a different notation (note that setting just 
$\bg=0$ in our Eq. (\ref{quasimain}) gives an even more general expression
that makes explicit the dependence of the mean density profile of a peak on the local tidal field).
It is important to stress that $\langle \delta(\bq) \rangle_{\rm pk}\neq \langle  \delta(\bq) | \bar{\delta}\rangle= \bar{\delta}\,\bar{\xi}(\bq)/\sigma_0^2$.
In words,
the conditional mean field\footnote{Also the scatter around it changes, see Eq. (7.9) in BBKS.} decreases more rapidly around a density peak with respect to a random point with the same $\bar{\delta}$. The
exact shape of the profile depends on the Hessian matrix of the density at the peak. This is a consequence of the fact that $\mH$ correlates with the density field as we have discussed in \S\ref{curv}.

The formalism to set up initial conditions for $N$-body simulations in the presence of peak constraints has been developed by \citet{VB}.  
This technique combines the HR method with the BBKS conditional probabilities, i.e. the
conditional mean field $\mu^{\rm (c)}(\bq)$ in Eq. (\ref{mfmulti}) is computed using expectations over
the point process formed by the density peaks $\langle \delta(\bq)| F_i[\delta]=f_i  \rangle_{\rm pk}$.
As the random realisation $\du$ does not have a peak at $\bq=0$, the final expression for the HR correction is
\begin{eqnarray}
\dc(\bq)-\du(\bq)\!\!\!\!\!&=&\!\!\!\!\!\langle \delta(\bq)| F_i[\delta]=f_i, \bar{\delta}, \bar{\bg}=0, \overline{\mH}\rangle\nonumber\\
&-&\!\!\!\!\!\langle \delta(\bq)| F_i[\delta]=f_{{\rm r},i}, \bar{\delta}_{\rm r}, \bar{\bg}_{\rm r}, \overline{\mH}_{\rm r}\rangle\;.
\end{eqnarray}
Eq. (\ref{peakcond}) shows that imposing the presence of a peak at a particular location requires specifying at least 10 constraints (1 for $\bar{\delta}$, 3 for $\bar{\bg}$ and 6 for $\overline{\mH}$) plus 
choosing a smoothing kernel and fixing its scale length.
However, additional requirements can be added.
For instance,
\citet{VB} also considered the linear velocity of the peak (or, equivalently, the
gravitational acceleration) and the linear velocity shear (or the traceless tidal tensor). In this case, there are
8 additional constraints to set. 
As in every other application of the HR method,
the constraints determine the conditional mean field and $\du$ provides the statistical noise around the
expectation.
By changing $\du$ for a given set of constraints, it is in principle possible to build an infinite number of realisations including all the large-scale environments that may exist. 
The method thus provides an unbiased sampling of the initial conditions that are compatible with the peak constraint.

Peak constraints are
particularly suitable for simulating the formation of structures that originate from rare field configurations. In fact these initial
conditions would be hardly encountered in random realisations of $\delta$.
Among the applications of the method are high-redshift quasars \citep[e.g.][]{RD11} and galaxy 
clusters \citep[e.g.][]{Dom06} as well as theoretical studies of gravitational collapse \citep[e.g][]{vdWB}.

\section{Genetically modified haloes} \label{gmh}
RPP applied the HR method to modify the initial conditions of $N$-body simulations within the Lagrangian patches that lead
to the formation of specific haloes (that, in the authors' jargon, get genetically modified, hereafter GM). The gist of the paper is to produce halo families in which the mass accretion history varies in a controlled and nearly continuous way.

In practice, the proposed method for genetic modification 
consists of several steps: 
i) a reference $N$-body simulation is run
starting from random initial conditions (i.e. from an unconstrained
realisation of a Gaussian field);
ii) a particular dark-matter halo is selected; 
iii) linear constraints are imposed (using the HR method) within the Lagrangian volume occupied by the 
particles that form the halo in the reference simulation; 
iv) a new simulation is run starting from the constrained initial conditions.

Genetic modification has a distinctive characteristic when 
compared with other applications of the HR method. In fact, 
it does not use statistical sampling: given a set of constraints,
there is one and only one realisation satisfying them. 
In a sense, the goal is to keep the large-scale structure fixed while altering
the linear density field around protohaloes and within a few correlation lengths of the variables on which the constraints are imposed. 
This objective could also be achieved by setting peak constraints as discussed in \S\ref{hrbbks} and smoothly changing the characteristics of the imposed peak (or enforcing simultaneous peak constraints on different length scales) while keeping $\du$ fixed. In compact notation, the peak-based analogue of genetic modification would be
\begin{equation}
\delta_{\rm pk2}(\bq)-\delta_{\rm pk1}(\bq)=\langle \delta(\bq)\rangle_{\rm pk2}-\langle \delta(\bq)\rangle_{\rm pk1}\;,
\label{peakgm}
\end{equation}
where both $\delta_{\rm pk1}$ and $\delta_{\rm pk2}$ are obtained from the same $\du$.
However, a strong point in favour of genetic modification is that it deals directly with protohaloes 
and does not rely on the assumption that virialised structures
form out of density peaks. In this Section, we are going to demonstrate that this advantage in theory turns out to be also a serious disadvantage in practical applications. Since we cannot yet associate protohaloes (and the characteristics of the corresponding haloes) with
particular configurations in the underlying density field, genetic-modification schemes currently have to trade exactness for tractability. 
Related to this,
we are going to show that the original implementation of
the genetic-modification algorithm by RPP is based on an unstated simplifying assumption and is therefore not exact
but approximate.
The degree of inaccuracy caused by this issue (in terms of the final halo structure and the mass accretion history)
is, however, difficult to gauge because of the highly non-linear dynamics of gravitational collapse. 
In this Section, we will focus on the conceptual issues while we will discuss practicalities in Section \ref{secmah}.

\subsection{Conditional averages at protohaloes}
\label{caap}
A key feature of the classic HR method is that the unconstrained field $\du$ is only used to generate
the noise around the conditional mean field. All the localised constraints are imposed
at random positions (e.g. at points with fixed coordinates)
for different realisations of $\du$ and `know' nothing about $\du$.
On the other hand, in order to implement their scheme for genetic modification, RPP
use information extracted from $\du$ to select
the location at which the constraints are imposed (as well as the shape and size of
the smoothing volume used to define the constraints). 
Genetic modification aims at transforming the Lagrangian regions of haloes. 
Therefore, not only the constraints are set only where $\du$ displays particular features, but 
it also is necessary that $\dc$ presents all the special features that define a protohalo at
the very same locations. 
From the mathematical point of view, restricting the analysis to protohaloes corresponds to changing 
the ensemble over which the conditional mean fields in Eqs. (\ref{mfmulti}) and (\ref{HRmult})
should be evaluated. 
Specifically, expectations should be taken over the point process formed by the protohaloes, $\langle \delta(\bq)| F_i[\delta]=f_i \rangle_{\rm h}$, although these are problematic to compute in practice.
This subtlety has been disregarded by RPP who instead derived the conditional mean by averaging over the distribution 
of the underlying overdensity field, $\langle \delta(\bq)| F_i[\delta]=f_i \rangle$, which is easy to work out.
Generally, this simplification introduces a bias in the constrained field, as we will show in detail later.
In summary, a self-consistent genetic-modification scheme should replace Eq. (\ref{HRmult}) with
\begin{equation}
\dc(\bq)-\du(\bq)=\langle \delta(\bq)| F_i[\delta]=f_{{\rm c},i}\rangle_{\rm h_c}-\langle \delta(\bq)|F_i[\delta]=f_{{\rm r},i}\rangle_{\rm h_r}\;,
\label{gmgm}
\end{equation}
in which the subscripts ${\rm h_c}$ and ${\rm h_r}$ distinguish the attributes of the different protohaloes.
Note that this expression closely parallels Eq. (\ref{peakgm}).

The main problem for integrating the HR method into the genetic-modification scheme concerns the identification of the protohalo sites and the statistical properties of the linear density field at these special locations.
In $N$-body simulations, protohaloes appear to be mostly associated with local maxima of the smoothed linear density field \citep{LP, HP14}. This tight correspondence is expected to produce a very specific form of scale-dependent bias between the clustering properties of the protohaloes and the underlying matter distribution \citep[BBKS,][]{Desjacques08} which is robustly measured in numerical simulations
\citep{ELP, Baldauf15}.
Simulations also show that the shape and orientation of proto-haloes strongly align with the local tidal field
\citep{LeeP, PDH2, LHP09, LP, Despali13, LBP}.
All these phenomena establish a link between the collapsing patches and 
several properties of the linear perturbations.
The emerging picture is that 
 the local values of the density, of its first and second spatial derivatives, and of the tidal field form the minimal set of variables that are necessary to characterise protohaloes.
 This conclusion forms the basis for 
 our discussion of conditional probabilities at protohaloes in the remainder of the paper.
 At this point, it is useful to recall that 
 setting simultaneous constraints on $\overline{\mD}, \bar{\bg}$ and $\overline{\mH}$ yields the
 conditional mean field and the $\Delta\chi^2$ function given
 in Eqs. (\ref{quasimain}) and (\ref{chimain}).

\subsection{A worked-out example}
In order to clarify the practical impact of the ensemble choice,  
we consider a simple representative example that has been already discussed by RPP 
and highlight the reasons for which their method is not exact. 
Suppose we want to genetically modify a halo by imposing a single density constraint $\dv=\dv_{\rm c}$. 
First of all, the Lagrangian patch that forms the selected halo in $\du$ must be used to define the smoothing volume appearing in Eq. (\ref{eq:defconst}).
Then, some version of the HR algorithm needs to be implemented. Starting from Eq. (\ref{meanHR}), RPP identify the mean-field correction
with the expectation of the density profile around random points having $\dv=\dv_{\rm c}$, i.e. ${\mu}^{\rm (c)}_{\bar{\delta}_{\rm c}}(\bq)=\langle \delta(\bq) | \,\dv=\dv_{\rm c} \rangle$
which leads to Eq. (\ref{sol1dens}).
This choice neglects that protohaloes form at special locations and treats them as any other point at which $\dv=\dv_{\rm c}$.
The ensemble average is blind to the value of either $\overline{\nabla \delta}$ or $\overline{\mH}$
(or even the tidal field) evaluated at the centre of the selected protohalo.
In fact, Eq. (\ref{sol1dens}) is obtained 
considering probability densities that have been
marginalised over all the field properties except the overdensity.
The resulting mean field would be meaningful if the value of $\dv$ would be the only information that matters to determine a protohalo.
However, this is not the case in general: protohalo sites are determined by additional field variables (see \S\ref{caap} for a plausible list).
Note that the HR method is exact. The inconsistency of the genetic-modification algorithm 
lies in the implicit assumption that protohaloes (where the constraints are set)
sample random points with a specific value of $\dv$. 
As we mentioned earlier, 
what one should do is to replace the conditional probabilities
$\langle \delta(\bq)| \,\dv=\dv_{\rm c}\rangle$ with 
$\langle \delta(\bq)| \,\dv=\dv_{\rm c}\rangle_{\rm h}$
where only the realisations that produce a protohalo at $\bq=0$ are considered in the ensemble 
average. Although this change provides the correct solution, we cannot evaluate the expectation value because
we do not know yet how to precisely characterise the locations of the protohaloes in mathematical terms. This is a formidable complication.

% Figure 2
\begin{figure*}
	\includegraphics[width=\columnwidth]{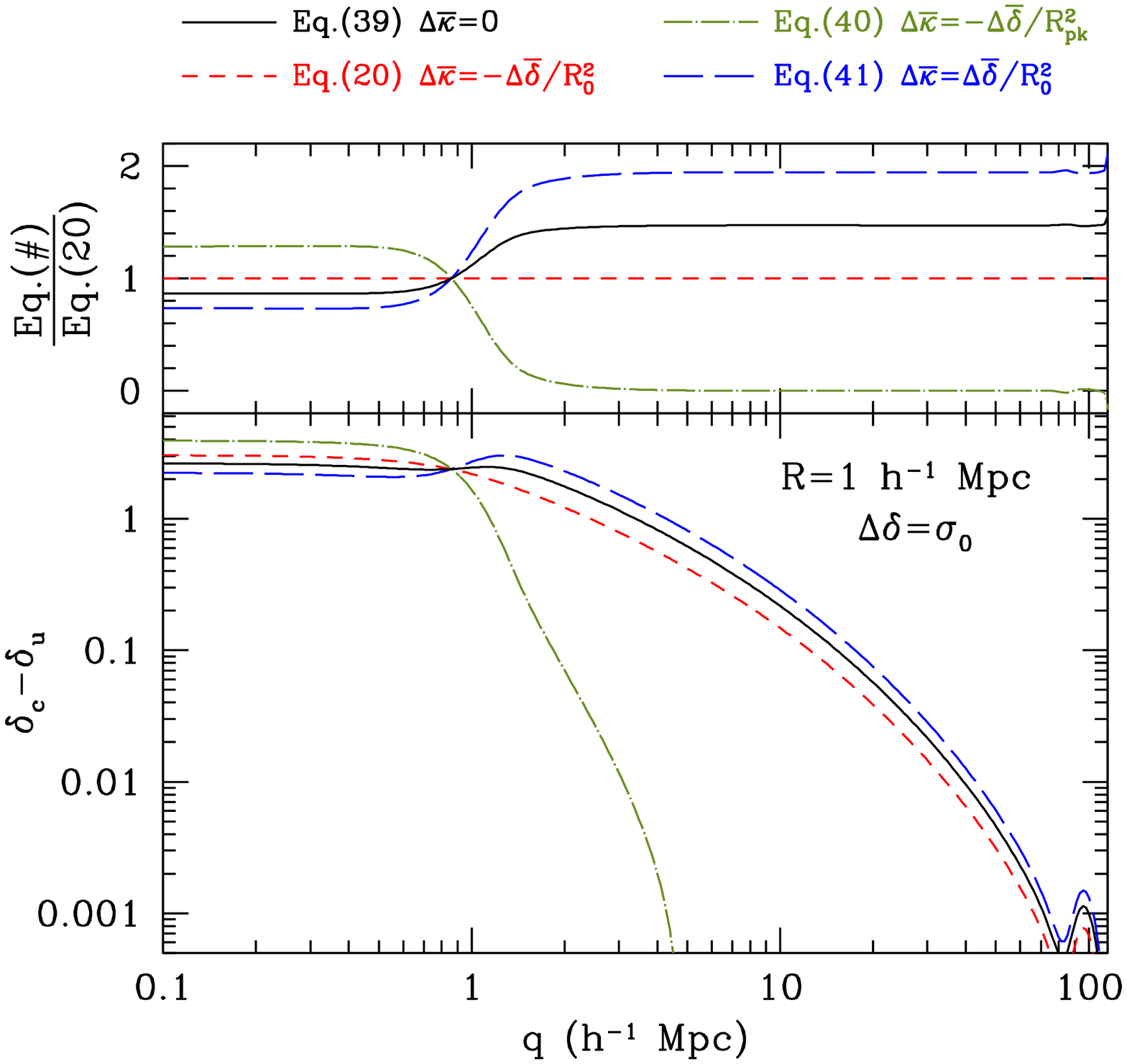}
	\includegraphics[width=\columnwidth]{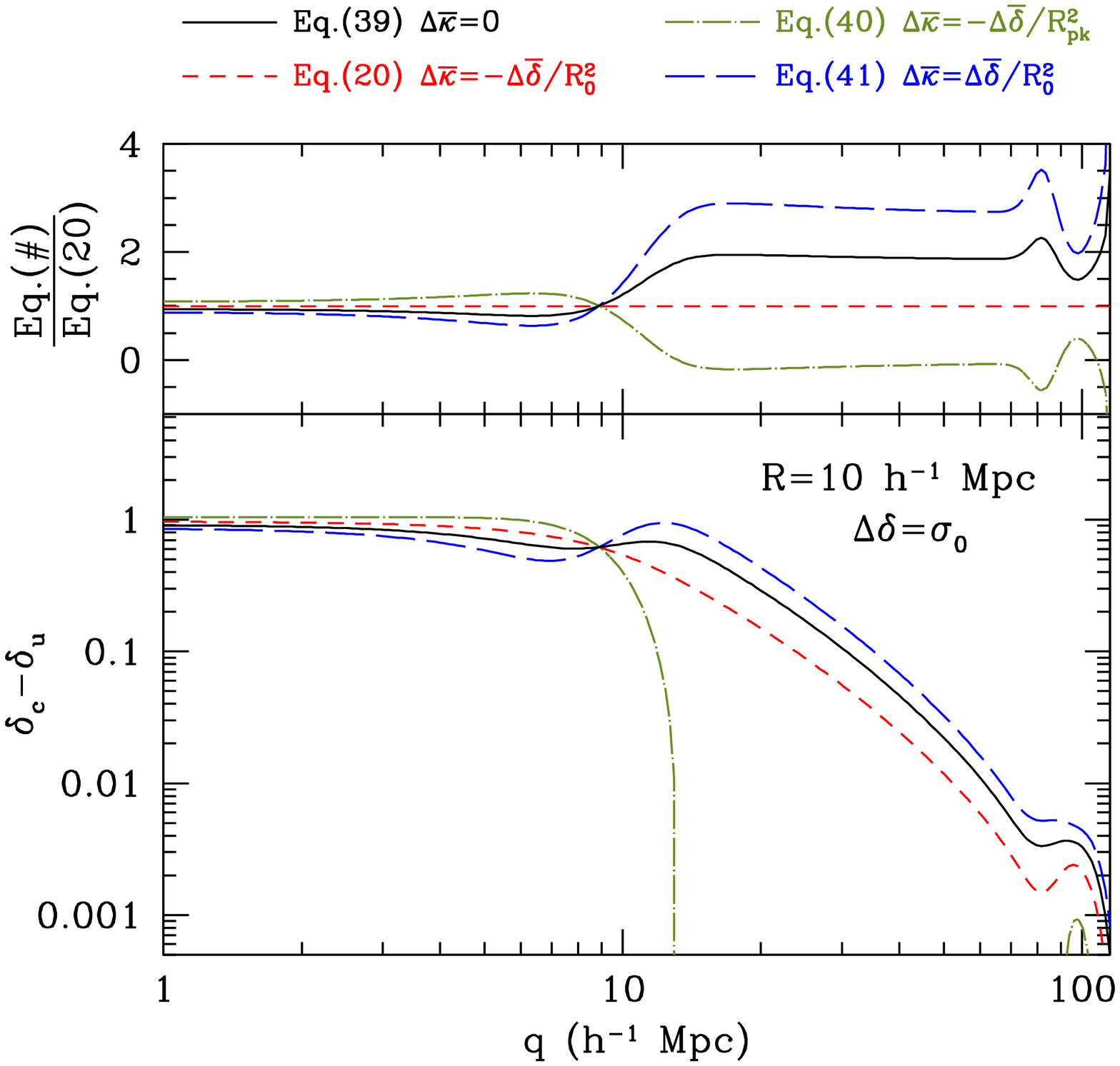}
    \caption{Field corrections $\dc-\du$ that generate the same change in the mean density ($\Delta\bar{\delta}=\sigma_0$) but different changes in the mean curvature within a region of Lagrangian size $R$. The left panel refers to galaxy-sized haloes ($R=1 \,h^{-1}$ Mpc) and the right
    panel to cluster-sized haloes ($R=10 \,h^{-1}$ Mpc).  Bottom: $\dc-\du$ vs. the distance from the protohalo centre. Top: ratio of the different functions with respect to the correction required to impose a density constraint at a random point, i.e. marginalizing over all the other field variables, used by RPP.}
    \label{fig2}
\end{figure*}

To better understand the problem, let us consider a couple of simpler 
cases for which we can write analytical solutions.
Let us assume for a moment that local extrema (i.e. maxima, minima and saddle points) of the linear
density field form a good proxy for the location of protohaloes.
By analogy with Eq. (\ref{peakcond}), the conditional probability enforcing $\bar{\delta}=\bar{\delta}_{\rm c}$ at an extremum can be written as (see Appendix \ref{condpeak})
\begin{equation}
\langle \delta(\bq) | \bar{\delta}=\bar{\delta}_{\rm c}\rangle_{\rm ex}=
\langle \delta(\bq)| \bar{\delta}=\bar{\delta}_{\rm c}, \bar{\bf s}=0\rangle\;.
\end{equation}
Taking into account the results presented in \S\ref{extre}, we thus
require that $\bar{\bg}_{{\rm c}}=0$ (i.e. the point at which the
constraints are set must be a density extremum in the constrained realisation)
and also assume that $\bar{\bg}_{{\rm r}}=0$ (i.e. the point was already
a density extremum in the unconstrained realisation).
In this case, from Eq. (\ref{HRdandg}) we indeed recover Eq. (\ref{sol1dens}) meaning that there is no difference in imposing density constraints at random points
or at density extrema with the same density. This happens because density and density gradients are uncorrelated. If haloes would form at density extrema, then the solution for setting constraints on $\bar{\delta}$ presented by RPP would be correct.

As a more realistic example, let us now assume that haloes form around linear density maxima of the $\delta$ field smoothed on the halo mass scale
\citep[an excellent approximation for massive haloes, see][]{LP}. 
In this case, when a density constraint is enforced, it also is necessary to impose that $\overline{\nabla \delta}=0$ (at $\bq=0$ both in $\dc$ and in $\du$)
and the Hessian matrix $\overline{\mH}$ is negative definite. 
It follows that the conditional mean field coincides with the peak density profile given in Eq. (\ref{peakprofile}).
Thus, even if one decides to keep all the elements of $\overline{\mH}$ unchanged, 
imposing a simple density constraint will require the following HR correction:
\begin{equation}
\label{eur}
\dc(\bq)-\du(\bq)=\frac{\Delta\bar{\delta}}{\sigma_0^2(1-\gamma^2)}\,\left[\bar{\xi}(\bq)+
R_{\rm pk}^2\,\nabla^2\bar{\xi}(\bq)\right]\;,
\end{equation}
which deviates from the one used in RPP and, as a matter of fact, will generate a different mass accretion history for the GM haloes.
The mismatch derives from the fact that $\bar{\delta}$ and $\overline{\mH}$ are correlated variables and by selecting
density peaks we are implicitly setting a constraint on $\overline{\mH}$.

As we mentioned before,
numerical simulations suggest that $\overline{\mH}$ plays a role in determining the location of protohaloes.
If this conjecture is true, then the HR correction for genetic modification
will also depart from the RPP solution.
This can be easily understood following a different line of reasoning:
if we want to preserve the density gradient, the Hessian matrix, and the tidal field at a given point (not necessarily a local maximum) while changing the overdensity, Eq. (\ref{main}) reduces to Eq. (\ref{eur})
with $\Delta \chi^2=(\bar{\delta}_{\rm c}^2-\bar{\delta}_{\rm r}^2)/[\sigma_0^2\,(1-\gamma^2)]$.

As expected, requiring that density maxima are genetically modified into density maxima with similar characteristics (or, more in general, that the Hessian matrix at the location of the constraints is not changed)
provides a different field correction with respect to enforcing a density constraint at a random point as in RPP
(see Figure \ref{fig2}).
Also the associated $\Delta \chi^2$ changes (see Figure \ref{figchi}). 

\subsection{Setting density and curvature constraints}
\label{pks}
\begin{figure}
	\includegraphics[width=\columnwidth]{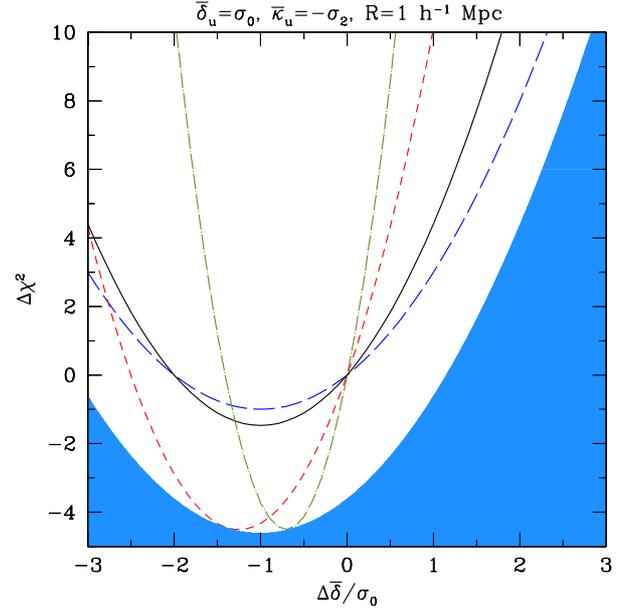}
    \caption{The relative likelihood of the fields $\dc$ and $\du$ is proportional to $\exp(-\Delta \chi^2/2)$.
    The quantity $\Delta\chi^2$ is plotted as a function of $\Delta \bar{\delta}$ for the different field transformations that have been presented in Figure \ref{fig2} (line styles are the same). Calculations are based on Eq. (\ref{chimain}). The shaded region indicates the values that cannot be obtained by imposing constraints that only change $\bar{\delta}$ and $\bar{\kappa}$ at $\bq=0$. Note that the field transformation given in Eq. (\ref{sol1dens}) is associated either with $\Delta\chi^2_{\rm ran}$ (long-dashed line) if interpreted as setting pure density constraints at random points (as in RPP) or with a different $\Delta\chi^2$ function (short-dashed line) if interpreted as setting joint constraints on $\bar{\delta}$ and $\bar{\kappa}$ with $\Delta\bar{\kappa}=-\Delta\bar{\delta}/R_0^2$.}
    \label{figchi}
\end{figure}

Further understanding can be gained through a study of the field transformations that change only the spherical parts of the tensors in Eq. (\ref{main}), i.e. $\bar{\delta}$ and $\bar{\kappa}$.
In this case, the most general HR correction consists of a linear superposition of terms proportional to $\bar{\xi}(\bq)$ and to $\nabla^2\bar{\xi}(\bq)$. The relative weight of the two contributions
depends on the exact form of the constraints. 
For instance, Eq. (\ref{main}) reduces to Eq. (\ref{sol1dens}) if $\Delta \bar{\kappa} =-\Delta \bar{\delta}/R_0^2$ while all the other variables are left unchanged. This means that what RPP call a `pure-density' constraint sets in reality correlated constraints on the density and the mean curvature\footnote{It is easy to understand how this works when we use a spherically symmetric filter: starting from the definition of $\bar{\xi}$ and
smoothing over the window function, one finds that $\bar{\bar{\xi}}({\bf 0})=\sigma_0^2$ and
$\overline{\nabla^2 \bar{\xi}}({\bf 0})=-\sigma_1^2$. Thus, using Eq. (\ref{sol1dens}) introduces the variation $\Delta \bar{\kappa}=-(\sigma_1^2/\sigma_0^2)\, \Delta\bar{\delta}=-\Delta \bar{\delta}/R_0^2$.} when one keeps $\bar{\bg}$, $\overline{T}_{ij}$ and $\overline{C}_{ij}$ fixed instead of marginalising over them (and $\bar{\kappa}$). 
In particular, if $\Delta\bar{\delta}<0$, the constraint can change sign to one or more of the principal curvatures and transform a density maximum into a saddle point or a minimum.
Moreover, the $\Delta \chi^2$ associated with the correlated constraints in the restricted ensemble is substantially different (see Figure \ref{figchi}) from what RPP found for random points, i.e. 
$\Delta \chi^2_{\rm ran}=(\bar{\delta}^2_{\rm c}-\bar{\delta}^2_{\rm r})/\sigma_0^2$. 
It is not surprising that the chance of drawing a specific realisation depends on the ensemble over which the probability has been defined: constraints that are likely in one ensemble might be rare in another one. In fact, the ensemble (i.e. what is kept fixed, what is marginalised over and what is allowed to vary) should always be specified when a quantity like $\Delta\chi^2$ is mentioned.

Another instructive example is obtained by requiring that $\Delta \bar{\kappa}=-\Delta \bar{\delta}/R_{\rm pk}^2$ which gives
\begin{equation}
\label{nablashape}
\dc(\bq)-\du(\bq)=\frac{\Delta\bar{\delta}}{\sigma_0^2(1-\gamma^2)}\,
(R_{\rm pk}^2-R_{0}^2)\,\nabla^2\bar{\xi}(\bq)\;.
\end{equation}
Note that imposing this constraint requires a field correction with a very different functional form than the previous ones (see Figure \ref{fig2}).
Finally, it is interesting  
to identify correlated constraints for which $\Delta\chi^2=\Delta\chi^2_{\rm ran}$. This is obtained imposing $\Delta\bar{\kappa}_{\rm c}=\Delta\bar{\delta}_{\rm c}/R_0^2$ which gives
\begin{equation}
\label{mostlik}
\dc(\bq)-\du(\bq)=\frac{\Delta\bar{\delta}}{\sigma_0^2(1-\gamma^2)}\,
(1+\gamma^2+2R_{\rm pk}^2\,\nabla^2)\,\bar{\xi}(\bq)\;.
\end{equation}
In Figure \ref{fig2}, we compare the expressions for $\dc-\du$ given in Eqs. (\ref{sol1dens}), (\ref{eur}), (\ref{nablashape}) and (\ref{mostlik})
assuming the spherically symmetric window given in Eq. (\ref{efffilter}) with two different smoothing radii, $R$.
All curves cross for $q$ slightly smaller than $R$ and their ordering is reversed for smaller and larger scales.
Moreover, since $\nabla^2\bar{\xi}$ drops much faster than $\bar{\xi}$ with increasing $q$, the field correction in Eq. (\ref{nablashape}) gives appreciable contributions only on scales comparable with $R$ or smaller and on the scale of the baryonic acoustic peak \citep[see also][]{Desjacques08}. On the other hand, all other expressions for $\dc-\du$ scale proportionally to  $\bar{\xi}$ on large scales but with substantially different normalisations. Eqs. (\ref{eur}) and (\ref{mostlik}) present a double peak (the first located at $q=0$ and the second for $q$ slightly above $R$) and have a positive slope at $q=R$.
These results show that, even only considering changes in the spherical parts of the deformation and density-Hessian tensors, there is quite some freedom in the choice of the constraints that fix the mean density within a protohalo in $\du$. Each transformation generates a different mass-accretion history
for the resulting halo and corresponds to a distinct protohalo shape in $\dc$.

In Figure \ref{figchi} we compare how the $\Delta\chi^2$ function varies with $\Delta \bar{\delta}$ for the different constrained fields considered so far.
Since $\Delta\chi^2$ also depends on the values that the functional constraints assume in $\du(\bq)$,
as a reference, we assume $\bar{\delta}_{\rm r}=\sigma_0$ and $\bar{\kappa}_{\rm r}=-\sigma_2$.
The boundary of the shaded region on the bottom indicates the lowest $\Delta\chi^2$ that can be obtained for a given $\Delta\bar{\delta}$ and is obtained minimising $\Delta\chi^2$ with respect to $\Delta \bar{\kappa}$ at fixed
$\Delta\bar{\delta}$. This corresponds to imposing $\bar{\kappa}_{\rm c}=\bar{\delta}_{\rm c}/R_0^2$.
The figure clearly illustrates that the relative likelihood of a constrained realisation does not depend only on the value of the density constraint (as assumed by RPP) but also on 
how the curvature of the perturbation is changed. Future applications of genetic modification should take this into account.

Before proceeding further, it is convenient 
to recap the main results presented in this Section.
First we have shown that
imposing constraints within the Lagrangian volume of haloes in 
the reference simulation (based on $\du$) introduces a bias due to the fact that the constraints
are implicitly set at special (i.e. non random) locations. This should 
be reflected in the conditional mean field of the HR formalism.  
Therefore, applying only a simple density constraint based on the statistics of random points as in RPP
is not conceptually rigorous and provides approximate results.
However, the state of the art does not allow us to provide a precise mathematical characterisation of protohaloes and thus
an exact algorithm for genetic modification cannot be formulated yet.
Using educated guesses based on the association between protohaloes and linear density maxima introduces several degrees of freedom into the
problem. 
In the next section, we will use the excursion-set model to quantify the actual
importance of this freedom in practical applications of the genetic-modification method and compare our results with
the original implementation by RPP.

\section{Predicting the mass accretion history}\label{secmah}
Changing at will the mass-accretion history of haloes by modifying the linear properties within the corresponding Lagrangian patches
would certainly be attractive and useful.
However,
the non-linear dynamics of halo collapse makes it difficult to predict the final outcome of the simulations given the initial constraints (or, vice versa, to pick the constraints that produce a given set of required properties).
RPP suggested that the final mass $M$ of the haloes forming from the constrained realisations can be accurately estimated using the halo mass function $n(M)$. Their key assumption is that   
the relative probability ${\cal P}_{\rm rel}$ of getting a perturbation with mean density $\dv$ coincides with the ratio $n(M)/n(M_{\rm r})$
where $M_{\rm r}$ denotes the mass of the halo formed in the unconstrained run.
RPP came up with a heuristic argument to test the consistency of this Ansatz when
only one density constraint is set (see their Section 6.1). 
Their reasoning assumes that there exists a well-defined $\dv$-$M$ relation and develops in terms of probabilistic arguments. 
It is difficult to understand, however, why the functional form of the halo mass function (which is a weighted average over all possible formation histories, i.e. over $\epsilon(\bq)$ and $\dv$) should be relevant for a problem which involves a single realisation of the residual field. 
Moreover, the HR method is completely deterministic (no generation of pseudo-random numbers is required to impose the constraints on a pre-existing random field) and this suggests that the relative probability of a constrained realisation should not matter at all to determine $M$.
In fact, for a given `family' of GM haloes - i.e. at fixed $\du(\bq)$ or $\epsilon(\bq)$ -  
there is a deterministic relation between $\dv$ and the final halo mass $M$ (even RPP approximated this relation with a power law for each
GM family). This relation, however, will be different for every realisation of the residual field. 
Similarly, the mass distribution within each family of GM haloes will depend on $\epsilon(\bq)$.
The mass function `emerges' only after averaging over the different realisations.

Here we use a variant of the excursion-set method in order to predict the
mass-accretion history and the final mass of the GM initial
conditions.

\subsection{Excursion sets}
\label{exc}
Let us consider a realisation of the linear density field and a specific halo that forms out of these initial conditions. 
The excursion-set trajectory, $\hat{\delta}(R)$, associated with the halo is obtained by averaging $\delta$ over a volume (with variable characteristic size $R$) surrounding the corresponding protohalo centre which we identify with the origin of the coordinate system. 
For instance, using a spherical top-hat filter $W_{\rm TH}(q)=3\,\Theta(R-q)/(4\pi R^3)$ with $\Theta(x)$ the Heaviside step distribution, one has
\begin{equation}
\hat{\delta}(R)=
\int W_{\rm TH}(q)\,\delta(\bq)\,\rd^3q
\end{equation}
(this is the same as in Eq. (\ref{eq:defconst}) but we will use $\bar{\delta}$ to indicate averages over the protohalo volume and $\hat{\delta}$ for averages over the excursion-set filter).
Depending on the application, the trajectory can be seen as a function of the smoothing radius $R$, the mass contained within the
filter in Lagrangian space $M=4\pi\bar{\rho}R^3/3$ (where $\bar{\rho}$ denotes the average comoving density of the universe) or the variance of the linear overdensity $\sigma_0^2$. 
It is convenient to sort the pseudo temporal variable in descending order for $R$ and $M$ and in ascending order for $\sigma_0^2$.
In what follows we will use $\log (M/{\rm M}_\odot)$.

% Figure 3
\begin{figure}
	\includegraphics[width=\columnwidth]{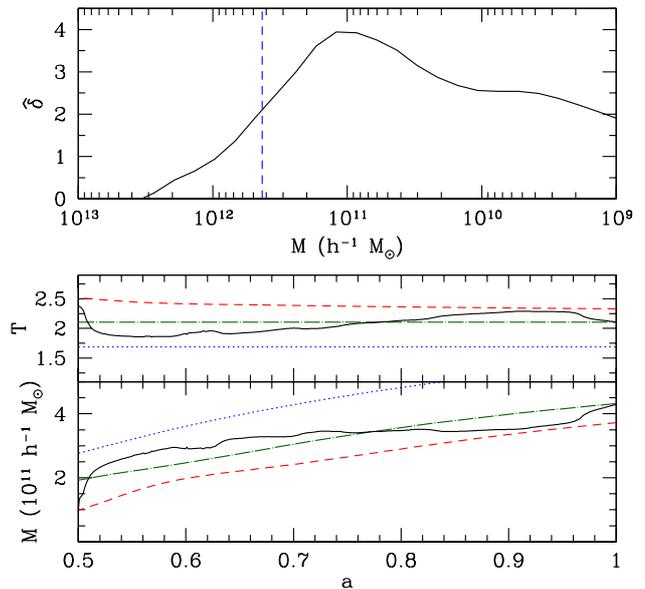}
    \caption{Top: Excursion-set trajectory centred on the Lagrangian region that forms a galaxy-sized halo in a high-resolution $N$-body 
    simulation (solid). The vertical dashed line indicates the halo mass at redshift $z=0$. Middle: The effective threshold $T$ which perfectly reproduces the mass accretion history of the halo (solid) is contrasted with the fit by \citet[][short dashed]{SMT01} and two constant
    thresholds: $T=1.686$ (dotted) and $T=2.1$ (dot-dashed).
Bottom: The mass-accretion history of the halo in the simulation (solid) is compared with the prediction of the excursion-set model using the thresholds shown in the middle panel. Choosing the constant value $T=2.1$ approximates the numerical data to better than 15 per cent.}
    \label{fig3}
\end{figure}

The excursion-set trajectory can be used to estimate the mass-accretion history of  every dark-matter halo \citep{BCEK}.
The key assumption is that the mass shell with Lagrangian radius $R$ will accrete onto the halo at time $t$ if 
$\hat{\delta}(R)$ - which scales with the linear growth factor $D_+(t)$ - is equal to a threshold $T$ and $\hat{\delta}(R')<T$ for all $R'>R$. Therefore, at a given epoch, the halo mass can be determined identifying the first upcrossing of the level $T$ by the excursion-set trajectory.
Detailed comparisons against $N$-body simulations have shown that this procedure works reasonably well if the trajectories are computed at protohalo centers while it fails miserably around random points \citep{White96, SMT01}.
The threshold value depends on the precise halo definition and several environmental factors that influence the geometry of gravitational collapse (e.g. the tidal field).
On average, it is a decreasing function of the halo mass but there is considerable scatter around the mean \citep{SMT01, Robertson09, ELP, LBP, BLP}.
Moreover, there exists a substantial population of low-mass haloes for which the excursion-set method works only at early times because tidal effects prevent the accretion of the outermost shells in Lagrangian space \citep{LBP, BLP}.

In the top panel of Figure \ref{fig3}, we show the excursion-set trajectory (linearly extrapolated at the present time, i.e. setting $D_+=1$) extracted from the initial conditions of a high-resolution $N$-body simulation and centered on the Lagrangian patch that forms a halo of mass $4.3 \times 10^{11}$ $h^{-1}$ M$_\odot$ at redshift $z=0$.
The halo has been identified using the AHF algorithm \citep{AHF} and the reported mass lies within a sphere with mean density $200 \rho_{\rm c}=200\bar{\rho}/\Omega_{\rm m}$ (here the matter density
parameter is $\Omega_{\rm m}=0.308$).
In the middle panel, we show the threshold value (solid) that would perfectly reproduce the mass-accretion history measured in the simulation.
For comparison, we also draw $T=1.686$ (dotted) as obtained from the collapse of a spherical top-hat perturbation in an Einstein-de Sitter universe and the mass-dependent fit derived by \citet[][short-dashed line]{SMT01}.
Note that the solid line lies always in between the other two. The fact that the effective threshold is larger than 1.686 is not surprising because tidal effects are expected to slow down gravitational collapse with respect to the spherical case. On the other hand, the threshold by \citet{SMT01} is statistical in nature as it
has been derived to fit the halo mass function and is not expected to accurately describe every single halo. 
The effective threshold that reproduces the $N$-body data oscillates around $T=2.1$ (dot-dashed) with relatively
small deviations (always smaller than 12 per cent). This constant threshold thus provides an excellent approximation for this halo between
$0\leq z\leq 1$. Note that at $z=1$ the halo undergoes a major merger and the point $\bq=0$ is contained in the Lagrangian region of the less massive progenitor. For this reason it 
it does not make sense to push the calculation for $z>1$.
Finally, in the bottom panel,
we contrast the mass-accretion history measured in the simulation (solid) with that predicted
by the excursion-set method using the different thresholds introduced above (same line styles as above).
The constant value $T=2.1$ reproduces the simulation masses to better than 15 per cent. For this reason we use this value in the remainder of the paper.
% Figure 4
\begin{figure}
		\includegraphics[width=\columnwidth]{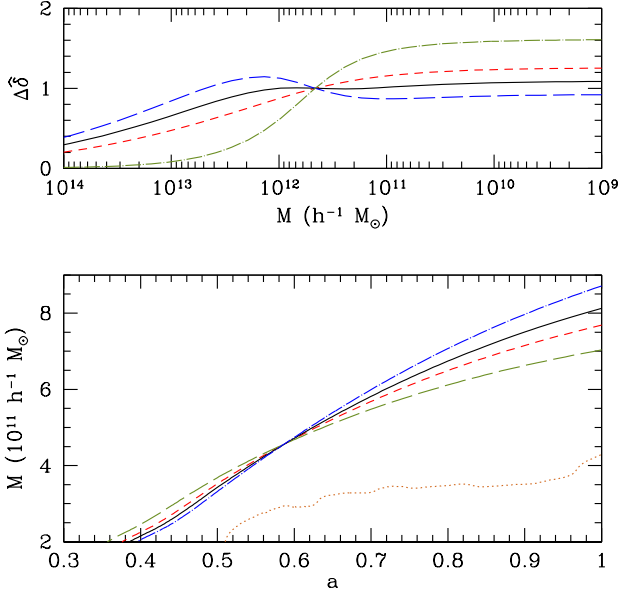}
    \caption{Top: Corrections to the excursion-set trajectory associated with  setting the constraint $\Delta \bar{\delta}=1$ 
    on the Lagrangian scale $R_{\rm c}=1\, h^{-1}$ Mpc. The line styles match those in Figure \ref{fig2} and refer to different
    constraints on the mean principal curvature at the protohalo centre.
    Bottom: Mass-accretion histories obtained applying the corrections shown in the top panel to the trajectory presented in Figure \ref{fig3}.  The excursion-set method with $T=2.1$ has been used to estimate the growth rate of the haloes stemming from the
    constrained realisations with $\Delta \bar{\delta}=1$.
As a reference, we also show the accretion history of the halo forming in the $N$-body simulation from the unconstrained initial
    conditions (dots).}
    \label{fig4}
\end{figure}

\subsection{Excursion set and genetically-modified haloes}
We now explain how the excursion-set method can be employed to predict the growth of GM haloes.
Let us first consider the simple density constraint presented  in Eq. (\ref{sol1dens}). The corresponding correction to the excursion-set trajectory is:
\begin{eqnarray}
\Delta \hat{\delta}(R)\!\!\!\!\!&=&\!\!\!\!\!
\hat{\delta}_{\rm c}(R)-\hat{\delta}_{\rm r}(R)=\frac{\Delta \bar{\delta}}{\sigma_0^2}\,\int W_{\rm TH}(q)\, \bar{\xi}(\bq)\,\rd^3\bq\nonumber\\
&=&\!\!\!\!\!\frac{\Delta \bar{\delta}}{\sigma_0^2}\,\int \widetilde{W}_{\rm TH}(kR)\,\widetilde{W}(\bk)\,P(k)\,\frac{\rd^3k}{(2\pi)^3}\\
&=&\!\!\!\!\!\Delta \bar{\delta} \, \frac{\langle \hat{\delta}(R)\, \bar{\delta}\rangle}{\langle \bar{\delta}^2\rangle}\nonumber
\end{eqnarray}
(this result follows from Eq. (\ref{xibarpk}) and the definition of Fourier transform).
Similarly, for the more complex case given in Eq. (\ref{eur}), one gets
\begin{eqnarray}
\Delta \hat{\delta}(R)\!\!\!\!\!&=&\!\!\!\!\!\frac{\Delta \bar{\delta}}{\sigma_0^2}\,\int \widetilde{W}_{\rm TH}(kR)\,\widetilde{W}(\bk)\,P(k)\,\frac{1-(kR_{\rm pk})^2}{1-\gamma^2}\,\frac{\rd^3k}{(2\pi)^3}\nonumber\\
&=&\!\!\!\!\!
\Delta \bar{\delta} \,
\frac{\sigma_2^2\,\langle  \hat{\delta}(R)\, \bar{\delta}\rangle^2-\sigma_1^2\, \langle  \hat{\delta}(R)\, \overline{\nabla^2\delta}\rangle}{\sigma_0^2\sigma_2^2(1-\gamma^2)}\;.
\end{eqnarray}
These corrections are completely deterministic and always the same independently of the unconstrained trajectory.
Consequently, there is no difficulty in computing $\hat{\delta}_{\rm c}(R)$.
To make a practical example, let us modify the initial conditions shown in Figure \ref{fig3} by requiring a density variation of $\Delta \bar{\delta}=1$ within a Lagrangian region of characteristic size $R_{\rm c}=1\,h^{-1}$ Mpc centered on the protohalo. We use  
Eqs. (\ref{sol1dens}), (\ref{eur}), (\ref{nablashape}) and (\ref{mostlik}) to set different correlated constraints on the mean curvature.
The resulting corrections to the trajectories\footnote{We use the window function in Eq. (\ref{efffilter}) to set the density constraint and a spherical top-hat filter to build the trajectories.} and the corresponding mass-accretion histories inferred from the excursion-set method are shown in Figure \ref{fig4}.
As expected, we find that the mass of the GM haloes assemble at a different rate depending on the exact form of the HR correction.
Our results clearly support two main conclusions. i) The excursion-set method provides a convenient tool to predict the non-linear
growth of the GM haloes. This procedure does not require any external input as the collapse threshold can be calibrated to match the mass-accretion history of the unconstrained realisation. 
ii) Although conceptually distinct, Eqs. (\ref{sol1dens}) and (\ref{eur}) generate similar 
mass accretion histories for galaxy-sized haloes although larger differences should be expected for cluster-sized haloes (see Figure \ref{fig2}).
This suggests that, after all,  the implementation by RPP might provide results in the right ball park, at least for certain classes of objects.
However, bigger discrepancies are found with Eqs. (\ref{nablashape}) and (\ref{mostlik}). The variable that appears to be most sensitive to the details
of the HR correction is the mass-accretion rate at the mass scale of the constraints (see below for a detailed explanation).
The extent to which the excursion-set method provides accurate predictions should be tested against $N$-body simulations, which is beyond the scope of this paper.

For constraints that induce relatively small changes in the trajectories, we can write an analytical expression for the mass variation. 
This is based on the fact that the slope of the trajectory determines how sensitive the final halo mass is to the
modifications induced by the constraints. Taylor expanding the unconstrained trajectory around the mass scale of first upcrossing
at a reference time $t_0$, we obtain
\begin{equation}
D_+(t_0)\,\hat{\delta}_{\rm r}(y)\simeq T+D_+(t_0)\,\hat{\delta}_{\rm r}'(y_{\rm up, r})\,(y-y_{\rm up, r})+\dots
\end{equation}
where $y=\log (M/{\rm M}_\odot)$ and $\hat{\delta}'=\rd \hat{\delta}/\rd y$ measures the slope (`velocity') of the excursion-set trajectory (note that this quantity is always negative at the scale of first upcrossing).
Similarly, assuming that the constraints are imposed at $y_{\rm up, r}$ (i.e. at the halo mass scale at time $t_0$), we get
\begin{equation}
\Delta\hat{\delta}(y)\simeq \Delta \bar{\delta}+ \Delta\hat{\delta}' (y_{\rm up, r})\,(y-y_{\rm up, r})+\dots\;.
\end{equation}
Finally, we can solve for the scale $y_{\rm up, c}$ at which $D_+(t)\,\hat{\delta}_{\rm c}(y_{\rm up, c})=T$ or, equivalently, for $\hat{\delta}_{\rm r}(y_{\rm up, c})+\Delta\hat{\delta}(y_{\rm up, c})=T/D_+(t)$ and find:
\begin{equation}
M_{\rm c}(t)=M_{\rm r}(t_0)\,10^{\alpha(t,t_0)}
\label{masspred}
\end{equation}
with
\begin{equation}
\alpha(t,t_0)=y_{\rm up, c}-y_{\rm up, r}=\frac{[T/D_+(t)]-[T/D_+(t_0)]-\Delta\bar{\delta}}{\hat{\delta}_{\rm r}'(y_{\rm up, r})+\Delta\hat{\delta}' (y_{\rm up, r})}\;.
\label{masspred2}
\end{equation}
For the halo in Figure \ref{fig3} this expression gives masses that are in very good agreement with those obtained using the full excursion-set model.

Eqs. (\ref{masspred}) and (\ref{masspred2}) acquire a particularly clear meaning for trajectories centred at local density maxima. In this case, 
the slope of the trajectory reflects the mean curvature of the peak \citep[this connection is remarkably transparent when Gaussian smoothing is used to build the trajectories, see also][]{Dalal08, MS12}.
The top panel of Figure \ref{fig4} shows that the sign of $\Delta\bar{\kappa}$ determines the slope of the corrections to the trajectory on the mass scale of the constraints and, consequently, the speed with which the halo mass grows.
Therefore, the freedom in setting simultaneous constraints on $\bar{\delta}$ and $\bar{\kappa}$ can be used to regulate both the final mass and the mass-accretion rate of the GM haloes. 

There are also other consequences of the curvature.
RPP have shown that different families of GM haloes occupy different loci in the plane defined by the concentration of the mass-density profiles and the collapse time (see the right panel in their Figure 4). Our discussion above provides new insight into the origin of this phenomenon. In fact, \citet{Dalal08} presented evidence from $N$-body simulations that steeper excursion-set trajectories correspond to haloes with higher mass concentration, at least for sufficiently large halo masses. Therefore, the offset in the tracks of the different GM families likely reflects the different slope of their excursion-set trajectories (i.e. the different curvature in the density at the protohalo location).

%% Example table
%\begin{table}
%	\centering
%	\caption{This is an example table. Captions appear above each table.
%	Remember to define the quantities, symbols and units used.}
%	\label{tab:example_table}
%	\begin{tabular}{lccr} % four columns, alignment for each
%		\hline
%		A & B & C & D\\
%		\hline
%		1 & 2 & 3 & 4\\
%		2 & 4 & 6 & 8\\
%		3 & 5 & 7 & 9\\
%		\hline
%	\end{tabular}
%\end{table}

\section{Angular-momentum constraints}\label{am}
RPP have pre-announced a forthcoming upgrade of their code in which they set constraints on the halo specific angular momentum.
In this Section, we extend our analysis to this type of constraints.
 
To leading order in the density and velocity perturbations, the angular momentum gained by a protohalo during its early-collapse phase is 
\citep{Doroshkevich70}
\begin{equation}
{\mathbf L}=-C\int W(\bq)\,\bq\times \nabla\Phi(\bq)\,\rd^3q
\label{L}
\end{equation}
where $\bq$ is measured from the centre of the protohalo and  $C$ is a time-dependent factor that follows from the fact that 
both the linear displacement of the mass elements and their linear velocity field are proportional to $-\nabla\Phi$.
Both $\mathbf{L}$ and the specific angular momentum per unit mass $\mathbf{L}/M$ (with $M=\bar{\rho}\,\int W(\bq)\,\rd^3q$)  
are thus linear in the density perturbations (as they scale proportionally to the peculiar potential) and suitable for the HR and the genetic-modification methods.
However, the angular momentum influences the process of gravitational collapse so that altering ${\mathbf L}$ necessarily changes the shape and size of the collapsing material and thus $W(\bq)$ in an unpredictable way.  For this reason, even ignoring the higher-order corrections to Eq. (\ref{L}), it is not possible to set precise constraints on the angular momentum (specific or not) of a GM halo.
What can be easily constrained, instead, is the linear angular momentum gained by a fixed Lagrangian volume corresponding to the window function $W(\bq)$, for instance the protohalo in the unconstrained initial conditions. 

Under the assumption made by RPP that protohaloes sample random locations with a given overdensity,
constraints on the Cartesian components of $\mathbf{L}$ can be easily imposed using Eq. (\ref{HRmult}).
In this case, there are four scalar constraints $\bar{\delta}=\bar{\delta}_{\rm c}$ and $\mathbf{L}=\mathbf{L}_{\rm c}$
so that the covariance matrix of their functional forms is composed of the blocks
$\langle \mathbf{L} \mathbf{L} \rangle$, $\langle\bar{\delta}^2\rangle=\sigma_0^2$ and $\langle \mathbf{L}\,\bar{\delta}\rangle=0$ (because $\mathbf{L}\propto \nabla\Phi$ while $\delta\propto \nabla^2\Phi$ and $\Phi$ is a Gaussian random field). Therefore, constraints on the mean density 
within the window function are statistically independent of those on $\mathbf{L}$.
Finally, since  $\delta$ is stationary, we obtain\footnote{Eq. (\ref{LL}) follows from the fact that all 2-point correlators are completely determined by the scalar distance between the points. In fact, $\nabla f(r)= \hat{\br}\, \partial f/\partial r$ for a generic function $f$ that depends only on the radial coordinate.} 
\begin{equation}
\frac{\langle \mathbf{L}\, \mathbf{L}\rangle}{C^2}=
\int  W(\bx)\,W(\by)\, \frac{(\by\times\bx)\,(\by\times\bx)}{|\bx-\by |^2}\,\psi(|\bx-\by|)\,\rd^3x\,\rd^3y\;,
\label{LL}
\end{equation}
where
$\psi(r)=\partial^2 \xi_{\Phi}/\partial r^2$
and $\xi_{\Phi}=\nabla^{-4}\xi(r)$ denotes the autocovariance function of the potential $\Phi$.
This expression completes the calculation of the matrix $\mA$ in Eq. (\ref{HRmult}).
On the other hand,
the shape of the mean field in the presence of the constraints
is determined by the cross-covariance function between $\delta$ and $\mathbf{L}$
(as before, we denote the location at which the constraints are set with the coordinates $\bq=0$),
\begin{equation}
\langle \delta(\bq)\,{\mathbf L}\rangle=
-C\int W(\bx) \,\frac{\bx\times\bq}{|\bx-\bq|}\,\omega_1(|\bx-\bq|)\,\rd^3x\;,
\label{Ldelta}
\end{equation}
with
$\omega_n(r)=\partial^n [\nabla^{-2}\xi(r)]/\partial r^n$
where $\nabla^{-2}\xi$ is the cross-covariance function between $\delta$ and $\Phi$.
Putting everything together, the HR method gives:
\begin{equation}
\label{amrandom}
\dc(\bq)-\du(\bq)=\langle \delta(\bq)\,L_i\rangle\,\left(\langle \mathbf{L} \,\mathbf{L} \rangle^{-1}\right)_{ij}\,\Delta L_j
+\frac{\bar{\xi}(\bq)}{\sigma_0^2}\,\Delta\bar{\delta}
\;.
\end{equation}
This expression can be used to set simultaneous constraints on ${\mathbf L}$ and $\bar{\delta}$ within a fixed Lagrangian volume centred on a random point.

The linear angular-momentum also correlates with the $n$th-order spatial derivatives of $\delta$:
\begin{eqnarray}
\frac{\langle L_i\, \partial_j \dots \partial_\ell\delta(\bq) \rangle}{C}\!\!\!\!\!&=&\!\!\!\!\!-
\int W(\bx) \,
\frac{(\bx\times\bq)_i\,(\bx-\bq)_j\dots(\bx-\bq)_\ell}{|\bx-\bq |^{n+1}} \nonumber \\
& & \omega_{n+1}(|\bx-\bq|)\,\rd^3x\;.
\label{Lder}
\end{eqnarray}
Note that the cross-covariances in Eqs. (\ref{Ldelta}) and (\ref{Lder}) vanish 
for $\bq=0$ implying that angular-momentum constraints are 
independent from the (unfiltered) values of
the density, the density gradient, and the curvature matrix at the protohalo centre.  This follows from two facts: i) the angular momentum is measured with respect to the centre itself, and ii) the statistical isotropy of the density field combined with the cross product.
This does not mean, however, that Eq. (\ref{amrandom})
can also be used to set linear angular-momentum constraints at special locations (e.g. protohaloes or density maxima). 
As we have already discussed for the density constraints in Section \ref{gmh}, extra requirements must be set to make sure
that averages are taken at protohaloes and the full covariance matrix of the joint constraints needs to be inverted in this case.
For instance, in the peak approximation, 
$\langle \delta(\bq) | \mathbf{L} \rangle_{\rm pk}=\langle \delta(\bq) | \mathbf{L}, \bar{\delta}, \bar{\bg}=0,\overline{\mH}\rangle
\neq \langle \delta(\bq) | \mathbf{L}, \bar{\delta} \rangle$. 
In fact, while the cross-correlation coefficients $\langle \mathbf{L}\,\bar{\delta}\rangle$ and $\langle \mathbf{L}\,\overline{\mH}\rangle$ vanish
because they pair odd and even spatial derivatives of the Gaussian field $\Phi$,
the term $\langle \mathbf{L}\,\bar{\bg}\rangle$ does not. Actually,
\begin{equation}
\langle \mathbf{L}\,\bar{\bg}\rangle=-C\int W(\bx)\,W(\by)\,\bx\times \langle \nabla\Phi(\bx)\,\nabla\delta(\by)\rangle\, \rd^3x \,\rd^3y
\end{equation}
with $\int W(\by)\,\langle \nabla\Phi(\bx)\,\nabla\delta(\by)\rangle\,\rd^3y=-\nabla\nabla\nabla^{-2}\bar{\xi}(\bx-\by)$
(contrary to Eqs. (\ref{Ldelta}) and (\ref{Lder}) this expression cannot be simplified in terms of radial derivatives because, in general, $\bar{\xi}(\bq)$ is not isotropic due to the asphericity of the window function that defines a protohalo).
This implies that linear-angular-momentum constraints correlate with conditions imposed on the mean density gradient. In other words,
angular-momentum constraints set at extremal points of the density field require a different HR correction than for constraints set at
random points with the same overdensity.
The exact expression for the correction can be derived by inverting the covariance matrix of the constraints which is beyond the scope of this paper and can be more easily done numerically.

The expression for the linear angular momentum in Eq. (\ref{L}) can be simplified by assuming that only the large-scale modes of the potential
contribute. In this case one can smooth $\Phi$ over the protohalo and replace it with its Taylor expansion \citep{White84}.
The leading-order term is  
$L^{\rm(T)}_i\simeq C\,\epsilon_{ijk} {D}_{j\ell} (\bq=0)\,{Q}_{\ell k}$ 
where ${Q}_{ij}=\int W(\bq) \,q_i \,q_j \,\rd^3 q$ is the quadrupole moment of the protohalo.
Note that the spherical parts of $D_{ij}$ and ${Q}_{ij}$ do not contribute to the cross product and $\mathbf{L}$ can then be expressed in terms of the linear tidal tensor $T_{ij}(\bq=0)$ and the traceless quadrupole moment $Q_{ij}-(Q_{ii}/3)\,\delta_{ij}$.
This result
forms the heart of the so-called tidal-torque theory 
and is equivalent to assuming that the (linear) velocity shear is approximately constant within the protohalo.
This approximation gives unbiased angular momenta with respect to
Eq. (\ref{L})  but generates a scatter of $\sim30$ per cent in the amplitude and a characteristic deviation of $20-30$ degrees in the direction \citep{PDH1}.
Higher-order corrections couple mass multipole moments of order $n>2$ with $n$ spatial derivatives of
$\Phi$ \citep[see Eqs. (10) and (11) in][]{PDH1}.
To first order in this expansion and
for a fixed quadrupole tensor $Q_{ij}$ (corresponding to a fixed Lagragian patch), linear-angular-momentum constraints are therefore equivalent to 
constraints on the local value of the linear tidal tensor (or velocity shear) and can be set
using Eq. (\ref{main}) even at density peaks. 
Note that, 
in this case,
$\langle \delta(\bq)\,L^{\rm (T)}_i\rangle=C\,\epsilon_{ijk} Q_{\ell k}\, [\partial_j \partial_\ell \nabla^{-2} \bar{\xi}(\bq) ]$. 

\section{Summary and conclusions}\label{con}
The HR method provides an efficient tool to generate constrained realisations of Gaussian random fields in which certain
linear functionals of the field variables assume pre-defined values.
Although this technique has been around for 25 years, many researchers are not very familiar with it and still see it as arcane or esoteric.
Motivated by the intent to improve this situation,
in Section \ref{const}, we reviewed the basic principles of the HR method and made a number of examples for its application to cosmology, including peak-based constraints. 
We hope that our analytical results will provide a useful reference and help revealing the intrinsic simplicity of the algorithm. 

In Section \ref{gmh}, we discussed `genetically modified' haloes. 
RPP applied the HR  algorithm to modify the initial conditions of $N$-body simulations within and around the regions that collapse to form
dark-matter haloes. The gist of their initiative is to alter the linear density field at will so that to produce haloes with a set of desired properties after the non-linear evolution. At first sight, this project might appear a relatively straightforward application of the HR
method. However, it contains a subtle complication: the points at which the constraints are applied are chosen after inspecting the
unconstrained realisation. They are the Lagrangian locations at which haloes form and they must preserve this property after being genetically modified. From the mathematical point of view, this is equivalent to restricting the ensemble over which averages in the HR method should be taken in order to build the conditional mean field. 
RPP have disregarded this issue and used averages taken over the full ensemble. 
In other words, they treated protohaloes as randomly selected points with a given overdensity in Lagrangian space. This implicit assumption made the calculation possible but the results that follow from it are likely to suffer from a statistical bias.
Our paper provides a first step towards understanding this issue.

What makes the problem so challenging is that we do not know yet how to characterize protohaloes in mathematical terms.
Although it is currently impossible to find an exact answer,  reasonable lines of attack have been presented in the literature.
Two common assumptions are that i) the Lagrangian sites for halo formation coincide with local density maxima of the smoothed density field
\citep[e.g.][BBKS]{Doroshkevich70,K84, PH} and ii) the boundaries of protohaloes correspond to isodensity surfaces \citep[e.g.][]{HP, CT}.
Detailed tests against $N$-body simulations give strong support to the validity of the first hypothesis, at least for haloes above the
characteristic collapsed mass at each epoch \citep{LP}. On the other hand, protohaloes' principal directions and shapes have been found to strongly correlate with the local tidal field rather than with the density distribution \citep{LeeP, PDH2, LHP09, LP, Despali13, LBP}.
All this suggests that it should be possible to characterize (at least to some extent) the properties of protohaloes in terms of the
following variables:  the density contrast, its first and second spatial derivatives, and the tidal field. 
Using the HR method we derived an analytical formula for setting simultaneous constraints on all these quantities.
Our result is given in Eqs. (\ref{quasimain}) and (\ref{main}) while Eq. (\ref{chimain}) can be used to evaluate the relative probability of the constrained realisations with respect to the original one.

If one wants to make sure that a protohalo in the unconstrained initial conditions, $\du$, remains a protohalo in the constrained linear density field, $\dc$, only some of the relevant field variable should be allowed
to vary while some others should be kept fixed. There is some freedom here.
For instance, one might want to require that a density peak in $\du$ stays a peak in $\dc$ (i.e. $\overline{\nabla \delta}=0$ and  $\overline{\mH}$ is negative definite).
With this in mind, we showed that the field transformation that sets a pure density constraint and marginalises over all the other field variables (Eq. (\ref{sol1dens}) which has been used by RPP) corresponds to setting correlated constraints in $\bar{\delta}$ and the mean curvature $\bar{\kappa}/3$ when the condition of being a local extremum and the traceless Hessian matrix are kept fixed. Although the expression of the HR correction is identical in these two cases, the likelihood of the constrained realisations is quite different. This demonstrates that $\Delta\chi^2$ values should be interpreted with care as they depend on the assumptions that are made on the nature of the constraints.
We also provided several additional examples including
the case in which a density constraint is imposed while keeping the density gradient, the Hessian matrix and
the tidal field fixed, Eq. (\ref{eur}). 

In the second part of the paper (Section \ref{secmah}) we have developed a variant of the excursion-set formalism in order to predict the mass-accretion history
of GM haloes. This is key to optimising the choice of the constraints that should be set in order to produce haloes with the desired properties after
their non-linear collapse. Our method does not require any external input and can be used with all sorts of constraints. Basically, we first 
compute the change in the excursion-set trajectory induced by the HR method and then solve for the first-upcrossing of a
threshold which has been calibrated using the mass-accretion history of the original unconstrained run. The entire algorithm is very simple
to code and essentially takes no time to run.
For constraints that require small changes we derived an analytical expression for the final halo mass which is given in Eqs. (\ref{masspred}) and
(\ref{masspred2}). 

Our analysis indicates that, after all,
the implementation by RPP generates halo mass accretion histories that are qualitatively similar to those obtained assuming a correspondence between protohaloes and local density maxima, at least on galaxy scales (see Figure \ref{fig4}). 
However, we found that the mass-accretion rate at the mass scale of the constraints is very sensitive to the detailed form of the imposed restrictions. This suggests that the method used by RPP might be suitable for investigating broad evolutionary scenarios but care should be taken when using it to make precise quantitative measurements.
Future studies should test our semi-analytic results against $N$-body simulations. In particular, they should measure how big of an effect is obtained when additional conditions on the density gradient, the Hessian matrix
and the tidal field are combined with the pure density constraints used by RPP.

Finally, in Section \ref{am},  we discussed the possibility of using the HR method to constrain the angular momentum that a
halo gains to leading order in perturbation theory.  
We concluded that this is impossible to achieve because the 
shape of protohaloes depends on the initial conditions in an unknown (and thus unpredictable) way. 
Nevertheless, the HR method can be used to set constraints based on the angular momentum gained by a fixed Lagrangian region. 
We derived the corresponding analytical solution for patches centered on random points with a fixed overdensity which is given in Eqs. (\ref{LL}), (\ref{Ldelta}) and (\ref{amrandom}).  
We also demonstrated that this solution does not hold true for density maxima or, more generally, when information on $\overline{\nabla \delta}$ is used to identify 
the location of the constraints (and thus, most likely, for protohaloes).
On the other hand, using the tidal-torque theory to first order, we reduced the angular-momentum constraints to tidal-field constraints that can more easily be imposed at  special locations identified using spatial derivatives of the density field.

In conclusion, we would like to express the wish that future investigations will focus more and more onto the problem of characterising 
the locations and properties of protohaloes.

\section*{Acknowledgements}
We warmly thank Nina Roth for discussions regarding the genetic-modification algorithm and
Yehuda Hoffman for suggestions that improved the presentation of our results.
 This work was partly funded by the German Research Foundation (DFG)
through the Cooperative Research Center TRR33 `The Dark Universe'.

%%%%%%%%%%%%%%%%%%%%%%%%%%%%%%%%%%%%%%%%%%%%%%%%%%

%%%%%%%%%%%%%%%%%%%% REFERENCES %%%%%%%%%%%%%%%%%%

%%%%%%%%%%%%%%%%%%%%%%%%%%%%%%%%%%%%%%%%%%%%%%%%%%

%%%%%%%%%%%%%%%%% APPENDICES %%%%%%%%%%%%%%%%%%%%%

\appendix

\section{Inverse covariance for the constraints}
\label{inversion}
We show here how to invert the 15-dimensional covariance matrix of the constraints discussed in \S\ref{curv}.
Since the density gradient is independent from all the other variables, we will consider only the deformation tensor and
the Hessian of the density for which $\langle \overline{D}_{ij} \,\overline{D}_{\ell m}\rangle= (\sigma_0^2 /15)\, S_{ij\ell m}$,
$\langle \overline{D}_{ij} \,\overline{H}_{\ell m}\rangle= (\sigma_1^2 /15) \,S_{ij\ell m}$ and
$\langle \overline{H}_{ij}\, \overline{H}_{\ell m}\rangle= (\sigma_2^2/15) \,S_{ij\ell m}$ with
$S_{ij\ell m}=\delta_{ij}\delta_{\ell m}+\delta_{i\ell}\delta_{jm}+\delta_{im}\delta_{\ell j}$.
If we organise the six independent elements of each tensor (say $\overline{D}_{ij}$) in the form of a vector with elements $(\overline{D}_{11}, \overline{D}_{22}, \overline{D}_{33}, \overline{D}_{12}, \overline{D}_{13}, \overline{D}_{23})$, then the 12-dimensional covariance matrix can be written as
\begin{equation}
\mA=\begin{pmatrix}
\mB_0 & \mB_1\\
\mB_1& \mB_2
\end{pmatrix}
\end{equation}
where $\mB_0=\sigma_0^2 \,\mM$, $\mB_1=\sigma_1^2 \,\mM$ and
$\mB_2=\sigma_2^2 \,\mM$ with
\begin{equation}
\mM=\frac{1}{15}
\begin{pmatrix}
3 & 1 & 1 & 0 & 0 & 0\\
1 & 3 & 1 & 0 & 0 & 0\\
1 & 1 & 3 & 0 & 0 & 0\\
0 & 0 & 0 & 1 & 0 & 0\\
0 & 0 & 0 & 0 & 1 & 0\\
0 & 0 & 0 & 0 & 0 & 1
\end{pmatrix}\;.
\end{equation}
The inverse covariance thus also has a block structure
\begin{equation}
\mC^{-1}=
\begin{pmatrix}
\mE& \mF\\
\mF & \mG
\end{pmatrix}
\end{equation}
with 
\begin{eqnarray}
\mE\!\!\!\!\!&=&\!\!\!\!\!(\mB_0-\mB_1\mB_2^{-1}\mB_1)^{-1}=\frac{1}{\sigma_0^2\,(1-\gamma^2)}\,\mM^{-1} \nonumber \\
\mF\!\!\!\!\!&=&\!\!\!\!\! -(\mB_0-\mB_1\mB_2^{-1}\mB_1)^{-1} \mB_1\mB_2^{-1}=- \frac{\gamma}{\sigma_0\,\sigma_2\,(1-\gamma^2)}\,\mM^{-1}\\
\mG\!\!\!\!\!&=&\!\!\!\!\! \mB_2^{-1}+\mB_2^{-1}\mB_1 (\mB_0-\mB_1\mB_2^{-1}\mB_1)^{-1} \mB_1\mB_2^{-1}=
\frac{1}{\sigma_2^2\,(1-\gamma^2)}\,\mM^{-1}\nonumber
\end{eqnarray}
where
\begin{equation}
\mM^{-1}=15
\begin{pmatrix}
2/5 & -1/10 & -1/10 & 0 & 0 & 0\\
-1/10 & 2/5 & -1/10 & 0 & 0 & 0\\
-1/10 & -1/10 & 2/5 & 0 & 0 & 0\\
0 & 0 & 0 & 1 & 0 & 0\\
0 & 0 & 0 & 0 & 1 & 0\\
0 & 0 & 0 & 0 & 0 & 1
\end{pmatrix}\;.
\end{equation}

\section{Conditional probabilities at local maxima}
\label{condpeak}
Let us consider a sufficiently smooth and differentiable Gaussian random field $\delta(\bq)$.
A mathematically well-defined cumulative probability distribution for the height of a local maximum of the field is obtained taking the limit\footnote{
Elementary probability theory cannot handle these probabilities because the event that a point process has an element
at a specified location has zero measure. Conditioning on point processes is rigorously defined in terms of the Palm distribution and Campbell measures \citep[see e.g.][]{DVJ07}. 
}
\begin{equation}
\lim_{\epsilon \to 0} {\cal{P}}\{ \delta(\bq_0)>u | \,\exists \mathrm{\ a\ local\ maximum\ of}\ \delta(\bq)\ \mathrm{in}\ U(\bq_0,\epsilon)\}
\end{equation}
where ${\cal P}$ denotes probability and $U(\bq_0,\epsilon)$ 
is the three-dimensional open cube of side $\epsilon$ centered at $\bq_0$ \citep{Cra-Lea}.
Generalising this definition to more variables and differentiating, we can introduce the differential probability distribution for maxima of height $\delta$ and (negative definite) Hessian matrix $\mH$,
$P_{\rm pk}(\delta,\mH)$, which, apart from a normalisation factor, coincides with the intensity function
$\bar{n}_{\rm pk}(\delta,\mH)$ such that $\bar{n}_{\rm pk}(\delta,\mH)\,\rd\delta\,\rd^6 \mH$ gives the expected number of peaks with height between $\delta$ and $\delta+\rd\delta$ and
Hessian matrix between $\mH$ and $\mH+\rd^6 \mH$ per unit comoving volume (note that $\rd^6 \mH$ denotes the Lebesgue measure on the space of $3\times 3$ negative definite matrices).

The intensity function
can be computed following the methods introduced by \citet{Kac} and \citet{Rice} as shown in BBKS. 
In brief, the reasoning proceeds as follows.
The number density of local maxima (characterised by the peak height $\delta_{\rm pk}$ and the Hessian matrix $\mH_{\rm pk}$) in one realisation of the random field can be formally written as
\begin{equation}
n_{\rm pk}(\bq)=\sum_i \delta_{\rm D}(\bq -\bq_{{\rm pk}, i})\;.
\label{formalnpk}
\end{equation}
Around a peak, the gradient of the random field can be approximated with its Taylor expansion to first order  $s_i(\bq)\simeq H_{ij}(\bq_{\rm pk})\,(\bq-\bq_{\rm pk})_j$.
Using the properties of the Dirac-$\delta$ distribution, Eq. (\ref{formalnpk}) can be re-written as
\begin{eqnarray}
n_{\rm pk}(\bq)\!\!\!\!\!&=&\!\!\!\!\!|\det{\mH}(\bq)|\,\{1-\Theta[\lambda_{\rm m}(\bq)]\}\nonumber\\ &\times&\!\!\!\!\!\delta_{\rm D}[\bg(\bq) ]\,\delta_{\rm D}[\delta(\bq)-\delta_{\rm pk}]\,\delta_{\rm D}[\mH(\bq)-\mH_{\rm pk}]\;,
\label{kacrice}
\end{eqnarray}
with $\lambda_{\rm m}$ the largest eigenvalue of $\mH$.
The function $\bar{n}_{\rm pk}$ is obtained taking the expectation of Eq. (\ref{kacrice}) which gives
\begin{equation}
\bar{n}_{\rm pk}(\delta_{\rm pk},\mH_{\rm pk})=
|\det{\mH_{\rm pk}} |\,[1-\Theta(\lambda_{\rm m, pk})]\,{\cal P}(\delta_{\rm pk}, {\bf s}=0, \mH_{\rm pk} )\;.
\end{equation}
where ${\cal P}$ is a multivariate Gaussian distribution expressing the joint probability of $\delta$, $\bg$ and $\mH$ in the original random field.

The conditional probability of a series of events $\mathbf{E}$ subject to the constraint that there is a density
peak at $\bq=0$ can be defined as the ratio between the number of peaks for which $\mathbf{E}$ is true and $\bar{n}_{\rm pk}$. The values assumed by the density field at all positions $\bq$ can also be included in $\mathbf{E}$.
Therefore, the conditional probability for a realisation of the field (here simply denoted by the letter $\delta$
and switching from functions to functionals), ${\cal P}_{\rm pk}[\delta | F_i[\delta]=f_i]$, can be formally written as
\begin{eqnarray}
%\langle \delta(\bq)| F[\delta]=f\rangle_{\rm pk}\!\!\!\!\!&=&
{\cal P}_{\rm pk}[\delta | F_i[\delta]=f_i]\!\!\!\!\!&=&
\!\!\!\!\!\frac{\bar{n}_{\rm pk}[\delta, \delta_{\rm pk}, \mH_{\rm pk}, F[\delta] =f]}{\bar{n}_{\rm pk}(\delta_{\rm pk}, \mH_{\rm pk})}\nonumber \\
&=&\!\!\!\!\!\frac{{\cal P}[ \delta, \delta_{\rm pk}, \bg=0, \mH_{\rm pk}, F[\delta] =f]}{{\cal P}(\delta_{\rm pk}, \bg=0, \mH_{\rm pk})}\nonumber\\
&=&\!\!\!\!\!\!{\cal P}[\delta |   \delta_{\rm pk}, \bg=0, \mH_{\rm pk}, F[\delta] =f]\;.
%&=&\!\!\!\!\!\langle \delta(\bq)| F[\delta]=f, \delta_{\rm pk}, \bg=0, \mH_{\rm pk} \rangle\;. 
\end{eqnarray}
In words, conditional probabilities at peaks coincide with conditional probabilities taken at random points characterised by $\delta=\delta_{\rm pk}$, $\bg=0$ and $\mH=\mH_{\rm pk}$.
It follows that the conditional mean field around a peak is
\begin{equation}
\langle \delta(\bq)| F_i[\delta]=f_i \rangle_{\rm pk}=\langle \delta(\bq)| F_i[\delta]=f_i, \delta_{\rm pk}, \bg=0, \mH_{\rm pk} \rangle\;.
\end{equation}
Similarly, for local extrema, one obtains: 
\begin{equation}
 \langle \delta(\bq)| F_i[\delta]=f_i \rangle_{\rm ex}=\langle \delta(\bq)| F_i[\delta]=f_i, \delta_{\rm ex}, \bg=0\rangle\;.
\end{equation}

% Don't change these lines
\bsp	% typesetting comment
\label{lastpage}
\end{document}